\def\v{{\bf v}}
\def\A{{\bf A}}
\def\matA{\undertilde{\A}}
\def\E{{\bf E}}
\def\B{{\bf B}}
\def\D{{\bf D}}

\def\j{{\bf j}}
\def\x{{\bf x}}
\def\p{{\bf p}}
\def\q{{\bf q}}
\def\k{{\bf k}}

\def\vecR{{\bf R}}
\def\R{{\rm R}}
\def\Wil{{\cal W}}
\def\O{\hat{\cal O}}
\def\Ok{\O(\k)}
\def\bzeta{\mbox{\boldmath$\zeta$}}
\def\grad{\mbox{\boldmath$\nabla$}}
\def\half{{\textstyle{1\over2}}}
\def\fourth{{\textstyle{1\over4}}}
\def\eighth{{\textstyle{1\over8}}}
\def\CA{C_{\rm A}}
\def\eps{\epsilon}
\def\gammaE{\gamma_{\rm E}}
\def\Tan{{\rm Tan}}
\def\tr{{\rm tr}}
\def\MSbar{$\overline{\hbox{MS}}$}

\def\IR{{\rm IR}}
\def\Re{{\rm Re}}
\def\Im{{\rm Im}}

\def\drangle{\rangle\!\rangle}
\def\dlangle{\langle\!\langle}

\def\Bigdrangle{\Bigr\rangle\!\Bigr\rangle}
\def\Bigdlangle{\Bigl\langle\!\Bigl\langle}
\def\Biggdrangle{\Biggr\rangle\!\!\Biggr\rangle}
\def\Biggdlangle{\Biggl\langle\!\!\Biggl\langle}

\def\deltaS{\delta^{S_2}}
\def\deltaC{\delta\hat C}
\def\deltac{\delta\hat c}
\def\Go{\hat G_0}
\def\Goo{\hat G_{00}}
\def\GGo{\hat{\cal G}_0}
\def\vevGo{\langle \Go \rangle}
\def\Gok{\hat G_0(\k)}
\def\Gop{\hat G_0(\p)}
\def\GGok{\hat{\cal G}_0(\k)}

\def\Po{{\hat P}_0}
\def\vA{\v\cdot\A}

\def\PL{P_{\rm L}}
\def\PT{P_{\rm T}}
\def\PiL{\Pi_{\rm L}}
\def\PiT{\Pi_{\rm T}}
\def\IT{I_{\rm T}}
\def\IL{I_{\rm L}}
\def\RT{R_{\rm T}}
\def\RL{R_{\rm L}}
\def\sig{\bar\sigma}
\def\sigo{\sig^{(0)}}
\def\sigmaL{\sigo_{\rm L}}
\def\sigmaT{\sigo_{\rm T}}
\def\rhoT{\rho_{\rm T}}
\def\rhoL{\rho_{\rm L}}
\def\ttot{t_\infty}
\def\eff{{\rm eff}}
\def\trans{\top}
\def\bare{{\rm bare}}
\def\dyn{{\rm dyn}}
\def\tot{{\rm tot}}

\def\bm{b^{(m)}}

\def\undertilde #1%
    {%
    \setbox0=\hbox {$#1$}%
    \setbox1=\hbox {$\tilde {\phantom {\copy0}}$}%
    \setbox2=\vtop {\offinterlineskip\box0\kern 1pt \box1}%
    \dp2 = 2.8 pt \box2\relax%
    }%

\def\wiggle{
   \mbox{
      \epsfig{file=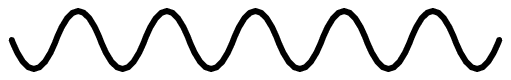,width=1in}
   }
}
\def\dashes{
   \raisebox{3pt}{\mbox{
      \epsfig{file=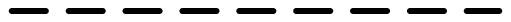,width=1in}}
   }
}

\def\underwick#1{{\vtop{\ialign{##\crcr
   $\hfil\displaystyle{#1}\hfil$\crcr\noalign{\kern3pt\nointerlineskip}
   \kern5pt\vrule\vbox to3pt{}\hrulefill\vrule\kern5pt
   \crcr\noalign{\kern3pt}}}}}
\def\overwick#1{{\vbox{\ialign{##\crcr\noalign{\kern3pt}
   \kern5pt\vrule\vtop to3pt{}\hrulefill\vrule\kern5pt
   \crcr\noalign{\kern3pt\nointerlineskip}
   $\hfil\displaystyle{#1}\hfil$\crcr}}}}

\def\underoverwick#1#2#3{\underwick{#1\overwickl{#2}}\overwickr{#3}}
\def\overwickl#1{{\vbox{\ialign{##\crcr\noalign{\kern3pt}
   \kern5pt\vrule\vtop to3pt{}\hrulefill
   \crcr\noalign{\kern3pt\nointerlineskip}
   $\hfil\displaystyle{#1}\hfil$\crcr}}}}
\def\overwickr#1{{\vbox{\ialign{##\crcr\noalign{\kern3pt}
   \vtop to3pt{}\hrulefill\vrule\kern5pt
   \crcr\noalign{\kern3pt\nointerlineskip}
   $\hfil\displaystyle{#1}\hfil$\crcr}}}}

\documentstyle[preprint,aps,eqsecnum,fixes,epsfig]{revtex}
\tightenlines
\begin {document}


\preprint {UW/PT 99--24}

\title
{High temperature color conductivity at next-to-leading log order}

\author {Peter Arnold}

\address
    {%
    Department of Physics,
    University of Virginia,
    Charlottesville, VA 22901
    }%
\author{Laurence G. Yaffe}
\address
    {%
    Department of Physics,
    University of Washington,
    Seattle, Washington 98195
    }%
\date {December 1999}

\maketitle
\vskip -20pt

\begin {abstract}%
{%
The non-Abelian analog of electrical conductivity at high temperature
has previously been known
only at leading logarithmic order:
that is, neglecting effects suppressed only by
an inverse logarithm of the gauge coupling.
We calculate the first sub-leading correction.
This has immediate application to improving, to next-to-leading log
order, both effective theories of non-perturbative color dynamics,
and calculations of the hot electroweak baryon number violation
rate.
}%
\end {abstract}

\thispagestyle{empty}


\section {Introduction}

We will provide a next-to-leading log order (NLLO) calculation of the
non-Abelian (or ``color'') conductivity in hot,
weakly-coupled non-Abelian plasmas.
The motivation for this calculation, and an overview of the strategy,
are presented in Ref.\ \cite{overview}.  Here, we will simply get down
to business.  ``Hot'' plasma means hot enough (1) to be ultra-relativistic,
(2) to ignore chemical potentials, (3) for non-Abelian gauge couplings
to be small, and (4) to be in the high-temperature symmetric phase if there
is a Higgs mechanism.

As discussed in Ref.\ \cite{overview}, there is a sequence of
effective theories which describe color dynamics at large distance
scales and long time scales.
\begin {mathletters}
\label {eq:1}
\begin {eqnarray}
\noalign {\hbox{\it Theory 1: $\omega,k \ll T$}}
&
   (D_t + \v\cdot\D)\, W - \v\cdot\E = 0 \,,
\label {eq:1a}
&\\&
   D_\nu F^{\mu\nu} = \jmath^\mu = m^2 \, \langle v^\mu W \rangle \,.
&
\label {eq:1b}
\end {eqnarray}
\end {mathletters}%
These ``hard-thermal-loop'' equations amount to linearized,
collisionless, non-Abelian, Boltzmann-Vlasov kinetic theory.
The parameter $m$ is the Debye screening mass,
which is $O(gT)$. 
This effective theory is valid, to leading order in the gauge coupling,
for frequencies and momenta small compared to the temperature,
$\omega,k \ll T$.

\goodbreak
\begin {mathletters}
\label {eq:2}
\begin {eqnarray}
\noalign {\hbox{\it Theory 2: $\omega \ll k \ll m$}}
&
   \v\cdot\D \, W - \v\cdot\E = - \deltaC \, W + \xi \,,
\label {eq:2a}
&\\&
   \langle W \rangle = 0 \,,
\label {eq:2b}
&\\&
   \D \times \B = \j  = m^2 \, \langle \v W \rangle \,,
&\\&
   \dlangle \xi \xi \drangle =
   {\displaystyle{2 T\over m^2} \, \delta C} \,.
&
\end {eqnarray}
\end {mathletters}%
This theory is a stochastic, collisional, linearized kinetic theory
of hard excitations coupled to slowly varying non-Abelian gauge fields.
Both the noise $\xi$, and the linearized collision operator $\delta \hat C$,
arise from integrating out the effects of gauge field fluctuations
below the scale of $m$.
Here, and henceforth, $\dlangle\cdots \drangle$
denotes an average over the (Gaussian) stochastic noise.
Theory 2 is valid for spatial momenta small compared to the
Debye screening mass and frequencies small compared to momenta,
$\omega \ll k \ll m = O(gT)$.
Eq. (\ref{eq:2b}) implements the effects of Gauss' Law in this
range of $\omega$ and $k$ and is explained in Ref.\ \cite{theory2}.

\begin {mathletters}
\label {eq:3}
\begin {eqnarray}
\noalign {\hbox{\it Theory 3: $\omega \ll k \ll \gamma$}}
&
   \sigma \E = \D \times \B + \bzeta \,,
\label {eq:3a}
&\\&
   \dlangle \bzeta \bzeta \drangle
       = 2 \sigma T \,.
\end {eqnarray}
\end {mathletters}%
This final theory is a stochastic Langevin equation,
known as B\"odeker's effective theory \cite{bodeker}
or ``the small frequency limit of Ampere's Law in a conductor'' \cite{Blog1}.
The parameter $\sigma$ is the ``color conductivity.''
Theory 3 is only valid on spatial momentum scales small compared to
the hard gluon damping rate $\gamma$
and frequencies small compared to momenta,
$\omega \ll k \ll \gamma = O[ g^2 T \ln(g^{-1}) ]$.
\smallskip

Our goal is to calculate, by successive matching of these effective
theories from short to large distance scales, the parameter $\sigma$
of Theory 3.

The spatial scale at which physics becomes non-perturbative is
$k \sim g^2 T$.  So, by at least a logarithm, the interfaces ($m$, $\gamma$)
of the successive effective theories are associated with perturbative physics,
thus making a perturbative matching calculation possible.
It will be useful to keep in mind a simple result from the analysis of
static properties of hot gauge theories, which is that the parameter which
controls the loop expansion is $g^2 T/k$, where $k$ is the momentum scale
of interest.
So, in particular, the loop expansion for physics at the interface
$k \sim \gamma$ between Theory 2 and Theory 3 is an expansion in
inverse logarithms $[\ln(1/g)]^{-1}$.

Our notation is the same as that of Ref.\ \cite{overview} and is
summarized in Table \ref{tab:notation}.
$W$ represents the adjoint color distribution of hard particles;
$\deltaC$ is a linearized collision operator;
$\xi$ and $\bzeta$ are Gaussian thermal noise;
$m$ is the leading-order Debye mass;
and $\gamma$ is the hard gluon damping rate.
Note in particular that $\langle\cdots\rangle$ denotes averages over
the direction $\v$ of hard particle velocities,%
\footnote{
   Readers of ref.\ \cite{BlaizotGamma} should beware that our use of the
   notation $\langle \cdots \rangle$ is completely different from theirs.
}
whereas
$\dlangle \cdots \drangle$ denotes averages over Gaussian noise.
Also note that we use $\delta C$ to represent both a $\v$-space integral
operator $\deltaC$ and the corresponding kernel $\delta C(\v,\v')$,
which is simply a function.
The formulas given earlier for the noise covariances
$\dlangle \xi \xi \drangle$ and $\dlangle \bzeta \bzeta \drangle$
should be understood as shorthand for

\begin {table}[t!]
\begin {center}
\begin {tabular}{l}
\hline
$v^\mu = (1,\v)$; $\v$ a spatial unit vector. \\
$\A = \A(\x,t)$, the spatial non-Abelian gauge field.  \\
$W = W(\x,\v,t)$,
    the adjoint color distribution of hard excitations. \\
$\bzeta = \bzeta(\x,t)$ and $\xi = \xi(\x,\v,t)$ are Gaussian white noise. \\
$\dlangle \cdots \drangle$ denotes averaging over noise. \\
$\langle \cdots \rangle \equiv \langle \cdots \rangle_\v$
    denotes averaging over the direction $\v$. \\
$\deltaS(\v{-}\v')$ is a $\delta$-function on the two-sphere
    normalized so that
   $\langle \deltaS(\v{-}\v')\rangle_{\v'} = 1$. \\
$\deltaC \, W \equiv \langle \delta C(\v,\v')\, W(\v') \rangle_{\v'}$,
    the linearized collision operator applied to $W$.\\
$\hat{}$ denotes either an operator on the space of functions of $\v$,
   or a spatial unit vector. \\
$\langle lm | \cdots | l'm' \rangle \equiv
   \int d\Omega_\v \> Y_{lm}^*(\v) \cdots Y_{lm}(\v)
   = 4\pi \langle Y_{lm}^*(\v) \cdots Y_{lm}(\v) \rangle_\v$. \\
$\langle l | \cdots | l' \rangle$ denotes the same for cases where
   the answer is proportional to $\delta_{m,m'}$. \\
$\Po \equiv {}|00\rangle\langle 00|$,
    the projection operator onto $\v$-independent functions. \\
$\gamma_1 \equiv \langle 1 | \deltaC | 1 \rangle$,
    the $l=1$ eigenvalue of the linearized collision operator.\\
$\CA$ is the adjoint Casimir of the gauge group [$N$ for SU($N$)]. \\
$d = 3-\eps$ with $\eps\to0$, the number of spatial dimensions. \\
$\displaystyle{\int_\q \equiv \int{d^dq\over(2\pi)^d}}$,
  and $\displaystyle{\int_{q_0} \equiv
       \int_{-\infty}^{+\infty}{dq_0\over2\pi}}$.\\
$(-+++)$ spacetime metric signature.\\
$\lambda \equiv q^0/q$, the ratio of frequency to spatial momenta. \\
$\approx$ denotes equality at leading-log order.\\
$\D = \grad + g\A^a T^a$, the gauge covariant derivative. \\
$T^b_{ac} = f^{abc}$, anti-Hermitian adjoint-representation generators.\\
``$\ln$'' in an order of magnitude estimate
    [{\em e.g.}\ $O(g^2 \ln)$] means $\ln(1/g)$.\\
\hline
\end {tabular}
\end {center}
\caption
   {%
   \label {tab:notation}
   Summary of notation.
   }%
\end {table}

\vspace*{1em}
\begin {eqnarray}
    \dlangle \xi^a(\v,\x,t) \, \xi^b(\v',\x',t') \drangle &=&
    {2 T\over m^2} \, \delta C(\v,\v') \, \delta^{ab}
    \> \delta^{(3)}(\x{-}\x') \> \delta(t{-}t') ,
\\
    \dlangle \zeta^a_i(\x,t) \, \zeta^b_j(\x',t') \drangle &=&
    2 \sigma T\,\delta_{ij} \, \delta^{ab}
    \> \delta^{(3)}(\x{-}\x') \> \delta(t{-}t') ,
\end {eqnarray}
where $i,j$ denote vector indices and $a,b$ are adjoint color indices.

The scale of the linearized collision operator $\deltaC$ is set
by $\gamma$.  At leading log order, it is given by \cite{bodeker,Blog1}
\begin {mathletters}
\label {eq:deltaC}
\begin {equation}
   \hat\delta C \, W(\v)
   \equiv \langle \delta C(\v,\v') \, W(\v') \rangle_{\v'} ,
\end {equation}
\begin {equation}
   \delta C(\v,\v') \approx
    \gamma \left[
       \deltaS(\v-\v')
       - {4\over\pi} \> {(\v\cdot\v')^2 \over \sqrt{1-(\v\cdot\v')^2}} 
    \right] .
\label {eq:deltaCleading}
\end {equation}
\end {mathletters}%
The symbol $\approx$ denotes equalities valid only to leading-log order.

We will consider the theories 1 and 2 to be ultraviolet (UV)
regulated by dimensional
regularization in $d=3{-}\eps$ dimensions with gauge
coupling $\mu^{\eps/2}g$.  Theory 3 is UV finite and does not
require such regularization.

In the remainder of this introduction, we review the path integral
formulation of effective theories 2 and 3, 
and summarize general properties of the collision operator
$\deltaC$ which we will need.
In section \ref{sec:sigma23}
we perform the matching of Theory 3 to Theory 2, which is
the most novel part of our calculation.
The NLLO conductivity $\sigma$
is calculated in terms of the collision operator $\deltaC$ of
Theory 2.
In section \ref{sec:sigma12}, we determine the information
we need about $\deltaC$ by matching Theory 2 to Theory 1.
In the process, we explicitly calculate the hard thermal gauge boson
damping rate in the presence of an infrared (IR) regulator
(specifically dimensional regularization).
Finally, in section \ref{sec:final}, we put everything together and
discuss the result.  We then also summarize the differences
between this work and an earlier discussion of the NLLO color
conductivity by Blaizot and Iancu \cite{BlaizotGamma}.



\subsection {Review of Path Integral Formulation}

The original derivation of the Langevin Eq.\ (\ref{eq:3})
by B\"odeker \cite{bodeker}, and
subsequent discussions \cite{Blog1}, were performed in $A_0=0$ gauge,
where the equation reads
\begin {equation}
   - \sigma {d\A\over dt} = \D\times\B + \bzeta \,
   .
\label {eq:A0}
\end {equation}
However,
$A_0=0$ gauge is a sick gauge for doing perturbation theory,
and is consequently an inappropriate choice for our present purposes.
It is therefore useful to reformulate (\ref{eq:A0}) as a path-integral,
so that we can use standard Faddeev-Popov methods to choose a more
convenient gauge.%
\footnote{
    The Langevin equation (\ref{eq:3}) may also be shown to be
    correct and unambiguous in general flow gauges of the form $A_0 = R[\A]$,
    where $R[\A]$ depends on $\A(\x,t)$ only
    instantaneously and so does not involve time derivatives of $\A$.
    See Ref.\ \cite{flow gauges} for a discussion of flow gauges,
    and Ref.\ \cite{zinnjustin&zwanziger} for a proof that
    the equation (\ref{eq:3}) may be applied in any gauge of this class.
    There are subtleties, however, in directly interpreting the Langevin
    equation (\ref{eq:3}) in {\it other}\/ gauge choices, such as Landau gauge.
    Using a flow gauge of the form $A_0 = \lambda \grad\cdot\A$
    (which is discussed further in appendix \ref {app:flow}),
    it should be feasible to reproduce all the analysis
    of this paper directly from the Langevin equation.
    However, we found it most straightforward,
    both conceptually and computationally,
    to use the path integral formulations presented here.
}
Eq.\ (\ref{eq:A0}) is
a Langevin equation, and it is well known how to reformulate such
equations as path integrals.%
\footnote{
   For a review, see chapters 4, 16, and 17, and in particular
   Eqs.\ (17.15) and (17.16), of Ref.\ \cite{ZinnJustin}.
   See also Ref.\ \cite{theory2}.
}
Specifically, Eq.\ (\ref{eq:A0}) becomes
\begin {mathletters}
\begin {equation}
   Z = \int [{\cal D}\A(\x,t)] \> \exp\left(-\int dt\>d^3 x \> L\right) ,
\end {equation}
with
\begin {equation}
   L = {1\over 4\sigma T}
          \left| \sigma \, {d\A\over dt} + \D\times\B \right|^2
     - \sigma^{-1} \, \delta^{(3)}(0) \> \tr \, \D^2
   ,
\label {eq:SA0}
\end {equation}
\end {mathletters}%
where $\D^2$ means $\D\cdot\D$.
We will use dimensional regularization throughout our analysis,
in which case
one may ignore the $\tr\, \D^2$ Jacobian term because
$\delta^{(d)}(0) \equiv 0$.
Eq.\ (\ref{eq:SA0})
is still in $A_0=0$ gauge, but we can now trivially generalize to
a gauge-invariant formulation:
\begin {mathletters}
\begin {equation}
   Z = \int [{\cal D}A_0(\x,t)] [{\cal D}\A(\x,t)] \>
         \exp\left(-\int dt\> d^3 x \> L\right) ,
\end {equation}
\begin {equation}
   L = {1\over 4\sigma T}
          \left| -\sigma\E + \D\times\B \right|^2 .
\label {eq:S3b}
\end {equation}%
\label {eq:S3}%
\end {mathletters}%
This can be checked by using the Faddeev-Popov procedure to return to
$A_0=0$ gauge.  But now we can use the Faddeev-Popov procedure on
(\ref{eq:S3}) to fix other gauges as well.
Coulomb gauge, for instance, corresponds to
\begin {mathletters}
\label {eq:Zcoulomb}
\begin {equation}
   Z_{\rm Coloumb} = \int [{\cal D}A_0] [{\cal D}\A]
          [{\cal D}\bar\eta] [{\cal D} \eta]
          \> \delta(\grad\cdot\A) \>
          \exp\left(-\int dt \> d^3 x \> L_{\rm Coloumb}\right) ,
\end {equation}
\begin {equation}
   L_{\rm Coloumb} = {1\over 4\sigma T} \left[
          \left| -\sigma\E + \D\times\B \right|^2
          + \bar\eta \grad\cdot\D \eta
   \right] ,
\label {eq:Lcoulomb}
\end {equation}
\end {mathletters}%
where $\bar\eta$ and $\eta$ are anti-commuting Faddeev-Popov ghosts.

Theory 2 can also be described by a path integral.  We shall find it
convenient to express the theory entirely in terms of the gauge fields
by eliminating $W$ using the equations of motion.
The resulting path integral formulation is discussed in detail in
Ref.\ \cite{theory2},
and here we simply quote the gauge-invariant result analogous to
(\ref{eq:S3}):%
\footnote{
   The path integral corresponding to this Lagrangian has time discretization
   ambiguities.  These should be resolved by a time-symmetric prescription.
   That is, writing $d\A/dt = [\A(t{+}\eps){-}\A(t)]/\eps$,
   and then interpreting
   $\A$'s without time derivatives to mean the average
   $[\A(t{+}\eps){+}\A(t)]/2$.
}
\begin {equation}
   L = {1\over 4T}
           \Bigl[ -\sig(\D)\, \E + \D\times\B \Bigr]^{\rm T}
           \sig(\D)^{-1}
           \Bigl[ -\sig(\D)\, \E + \D\times\B \Bigr]
    + L_1[\A],
\label {eq:L2}
\end {equation}
where $\sig(\D)$ is now a matrix in vector-index space.
It is also an operator in $\x$ space (and color)
and is given by
\begin {equation}
   \sig_{ij}(\D)
   = m^2 \lim_{\Lambda\to\infty} \Bigl\langle v_i \,
             [ \v \cdot\D + \deltaC + \Lambda \hat P_0 ]^{-1}
             v_j \Bigr\rangle
   =  m^2 \Bigl(
      \langle v_i \hat G v_j \rangle
      - \langle v_i \hat G \rangle \langle \hat G \rangle^{-1}
                    \langle \hat G v_j \rangle
   \Bigr) ,
\label{eq:sigmaD}
\end {equation}
where $\hat G$ is the $W$-field propagator arising from (\ref{eq:2a}):
\begin {equation}
   \hat G \equiv [\v\cdot\D + \deltaC]^{-1} .
\label{eq:Gdef}
\end {equation}
$\hat G$ is an operator in both $\x$ and $\v$ space
(as well as color).
In the first form of (\ref {eq:sigmaD}),
$\hat P_0$
denotes the $\v$-space projection operator that
projects out functions that are independent of $\v$.
In the notation introduced below, $\hat P_0 = |00\rangle\langle 00|$.

The term $L_1[\A]$ in (\ref{eq:L2}) is complicated and is discussed in
Ref.\ \cite{theory2}.
It is the analog of the $\delta^{(3)}(0) \, \tr \,\D^2$
term in (\ref{eq:SA0}) but is
spatially non-local and does not trivially vanish in dimensional
regularization.
Fortunately,
however, the size of $L_1[\A]$ is such that it will
be irrelevant to our calculation of the NLLO conductivity.
All we need to know about it for the present discussion
is that $L_1[\A]$ is independent of $A_0$,
and that it is suppressed by the loop expansion parameter compared
to other terms in (\ref{eq:L2}).
[Specifically, its contribution to the $\A$ propagator
is suppressed by one factor of the loop expansion parameter.]
As we will see later, this will be enough to argue that this
term does not affect our calculation of the conductivity at NLLO.

The trace of
$\E\cdot\D\times\B$ is the space-time total derivative
$-\grad \cdot (\E^a \times \B^a) - \partial_t \, (\half \B^a \B^a)$.
So, ignoring boundary terms, the cross-term may be dropped in the action
for (\ref{eq:L2}):%
\footnote{
   This reasoning depends on using a symmetric time discretization to
   define (\ref{eq:L2}).
   See, for example, Appendix B of Ref.\ \cite{langevin}.
   Time is to be regarded as running from $-\infty$ to $+\infty$,
   and the choice of temporal boundary condition is irrelevant
   for our purposes.
}
\begin {equation}
   S = \int dt \> d^3x \> L
     = \int dt \> d^3 x \left\{
        {1\over 4T}\left[
           \E \,\sig(\D) \E + (\D\times\B)\, \sig(\D)^{-1} (\D\times\B)
        \right]
        + g^2 L_1[\A]
     \right\} .
\label{eq:S2}
\end {equation}

We should perhaps clarify our mixture of
$\v$-space operator notation and
$\langle \cdots \rangle$ notation for averaging over $\v$.
Since $\deltaC$, $\hat P_0$, and $\hat G$ are operators in $\v$-space,
they do not commute with $v_i$ and $v_j$ in (\ref{eq:sigmaD}).
So, for instance,
$\langle v_i \, \deltaC \, v_i \rangle \not= \langle \deltaC \rangle$
even though $v_i v_i = |\v|^2 = 1$.
This particular example is made clear by rewriting
$\langle v_i \, \deltaC \, v_i \rangle
   = \langle v_i \, \delta C(\v,\v') \, v'_i \rangle_{\v\v'} .$
Our notation works just like bra-ket notation in quantum mechanics
if one rewrites $\langle \cdots \rangle$ as
$\langle 0 0 | \cdots | 0 0 \rangle$
with $|00\rangle$ representing the constant function
$Y_{00}(\v) = 1/\sqrt{4\pi}$ in
$\v$-space.
(It will be necessary later to consider other $|lm\rangle$
spherical harmonics as well.)
$|00\rangle$ is only a $\v$-space entity and does not specify anything
about $\x$ or color dependence.  So, for instance,
$\langle \D \rangle = \langle 00 | \D | 00 \rangle$ is $\D$ and not zero.


\subsection {General properties of $\deltaC$}
\label{sec:deltaC}

We'll now discuss some useful properties of $\deltaC$ that follow from
general considerations not restricted to the leading log approximation
(\ref{eq:deltaCleading}).  Collisions packaged in $\deltaC$ are, as far as
effective Theory 2
is concerned, local in space.  All of the $\x$ dependence of collision
probabilities comes in the distribution functions $W$, and so the operator
$\deltaC$ itself is independent of $\x$ --- it is simply an operator
in $\v$-space.  Rotation invariance of the theory
therefore implies $\v$-space rotation invariance of the operator $\delta C$,
which in turn implies that $\deltaC$ is diagonal in the space of
$|lm\rangle$'s.
[That is, its eigenfunctions are the spherical harmonics $Y_{lm}(\v)$.]
It also implies that the
corresponding diagonal matrix elements
$\langle lm | \deltaC |lm \rangle$ depend only on $l$ and not on $m$.
We shall therefore write them more compactly as
$\langle l | \deltaC | l \rangle$.
This $l$-decomposition of $\deltaC$ will be crucial to our
later analysis, and the case $l=1$ will be of particular interest.%
\footnote{
   For a discussion of using the space of spherical harmonics $|lm\rangle$
   as a basis for numerical simulations, see Ref.\ \cite{lsim}.
}

It's useful to demonstrate this notation by returning to the leading-log
approximation (\ref{eq:deltaCleading}).
The second term of (\ref{eq:deltaCleading})
vanishes when applied to any $\v$-parity-odd function, and so
$\langle l | \delta\hat C | l \rangle \approx \gamma$
for odd $l$.
In particular, at leading-log order,
\begin {equation}
   \langle 1 | \delta\hat C | 1 \rangle \approx \gamma .
\label {eq:C11leading}
\end {equation}
The relation is more complicated beyond leading-log
order, and we will define a separate symbol%
\footnote{
  In $d$ dimensions, $\gamma_1$ means the eigenvalue of $\delta\hat C$
  in the vector representation of SO($d$).
}
\begin {equation}
   \gamma_1 \equiv \langle 1 | \delta\hat C | 1 \rangle .
\label{eq:gamma1def}
\end {equation}
In section \ref{sec:sigma23}, we will find that the only pieces of
$\deltaC$ that we need to calculate the NLLO conductivity are the
leading-log formula (\ref{eq:C11leading}), plus the NLLO result for
$\gamma_1$.
The eigenvalue $\gamma_1$ will appear in our analysis through the handy
formulas
\begin {equation}
   \deltaC \, \v\rangle = \gamma_1 \v \rangle ,
   \qquad\qquad
   \langle \v \, \deltaC = \gamma_1 \langle \v .
\label{eq:dC1}
\end {equation}
Here, and henceforth, we will often use a bare $\rangle$
or $\langle$
as convenient shorthand for $|00\rangle$ or $\langle 00|$, respectively.
The relations (\ref {eq:dC1})
follow because $\v\rangle$ has $l=1$, and so
is a superposition of $|1m\rangle$'s.
The value of $\gamma_1$ will be calculated in section \ref{sec:sigma12}.

Rotation invariance also implies that $\deltaC$ is symmetric in $\v$-space,
since $\delta C(\v,\v')$ can only depend on $\v\cdot\v'$.
Furthermore, like everything else in Eq.~(\ref{eq:2a}), $\deltaC$ is real.
As discussed in Ref.\ \cite{theory2}, this means that the
conductivity operator $\sig^{ij}(\D)$ of (\ref{eq:sigmaD}) is real
and symmetric in $\x$/color/vector space.

A very important property of $\delta C$ is that it annihilates functions
which are independent of $\v$:
\begin {equation}
   \deltaC \rangle = 0 ,
\label{eq:dC0}
\end {equation}
or equivalently
\begin {equation}
   \langle \deltaC = 0 ,
   \qquad \mbox{or} \qquad
   \langle 0| \deltaC |0 \rangle = 0 .
\end {equation}
We will see this explicitly when we discuss $\deltaC$ in section
\ref{sec:sigma12}, but it is true for quite general reasons,
as pointed out in Refs.\ \cite{bodeker,Blog2}.
One way to understand this is to observe that current conservation
in effective theory 1 requires
$0 = D_\mu \jmath^\mu = D_\mu \langle v^\mu W \rangle$.
Theory 2 is a subsequent effective theory for $\omega \ll k$,
meaning that time derivatives have been neglected compared to
spatial derivatives.
In this limit, current conservation becomes simply
$\langle \v\cdot\D W \rangle = 0$.
Taking the $\v$-expectation value of (\ref{eq:2a}) gives
$\langle \v\cdot\D W \rangle = \langle \deltaC \rangle$, and so
$\langle 00 | \deltaC | 00 \rangle = 0$.
Since $\deltaC$ is diagonal in $|lm\rangle$ space, it follows that
$\deltaC$ has a zero-mode, $\deltaC \rangle = 0$.
A differently packaged but related explanation of this property
may be found in sec.\ III\ C of Ref.\ \cite{Blog2}.

Finally, the effect of a collision term in a Boltzmann equation is to
cause the decay over time of correlations.  The sign of $\deltaC$ in
(\ref{eq:2a}) is such that decay corresponds to positive $\deltaC$,
which would be more obvious if we hadn't dropped the $\partial_t W$ time
derivative term
in going from (\ref{eq:1a}) of Theory 1 to (\ref{eq:2a}) of Theory 2.
Based on this, one should expect that all the eigenvalues
$\langle l| \deltaC |l \rangle$ of $\deltaC$ are non-negative.
We shall verify this explicitly in section \ref{sec:SigFormula}
using the leading-log
formula (\ref{eq:deltaCleading}).
The zero eigenvalue (\ref{eq:dC0}) corresponds to the fact that charge
is conserved and does not decay.



\section {Matching Theory 2 to Theory 3}
\label {sec:sigma23}

To match effective theories, one must identify gauge-invariant
observables which are calculable in both theories.
We shall therefore spend some
time discussing a gauge-invariant observable, involving Wilson loops,
that can be used to determine the parameter $\sigma$ of Theory 3.
We shall then find that, in practice,
the matching problem can be conveniently simplified to the matching
of the Coulomb-gauge self-energy $\Pi_{00}$ of $A_0$
in the limit of zero frequency and small momentum.


\subsection {Wilson loops}
\label{sec:Wilson1}

\begin {figure}[t!]
\vbox{
   \begin {center}
      \epsfig{file=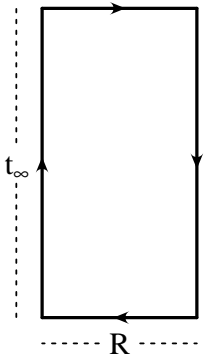,scale=.9}
   \end {center}
   \vspace*{-.2in}
   \caption{
       A time-like Wilson loop rectangle.
       \label{fig:Wilson0}
   }
}
\end {figure}

An example of a gauge-invariant observable that depends on the
conductivity is a (real-time) Wilson loop
\begin {equation}
   \Wil = \Biggdlangle {\rm tr}\>{\cal P} \>
   \exp\left(g \oint dx^\mu A_\mu\right)
       \Biggdrangle
\end {equation}
that extends in the time direction, where
${\cal P}$ indicates path ordering of the exponential.
Here $A_\mu \equiv A_\mu^a \, T^a$, and our convention is that the generators
$T^a$ are anti-Hermitian.
To see the dependence on conductivity explicitly, it is convenient to focus on
rectangular Wilson loops, such as depicted in Fig.\ \ref{fig:Wilson0}, where
one set of edges is in the time direction and the other set is
purely spatial.  It will also be convenient to focus on
rectangles whose temporal extent $\ttot$
is very large compared to their spatial extent $R$.

\subsubsection {Relation to $\sigma$ in Theory 3}

\begin {figure}[h!]
\vbox{
   \begin {center}
       \epsfig{file=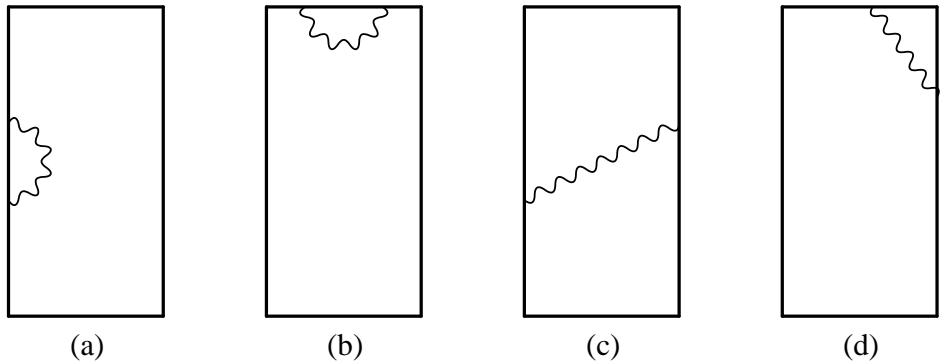,scale=.9}
   \end {center}
   \vspace*{-.2in}
   \caption{
       Examples of first-order contribution to the expectation of a
       large-time Wilson rectangle.
       \label{fig:Wilson1}
   }
}
\end {figure}

To get a feel for these Wilson loops, let's look at their value in
our final effective theory, Theory 3, at
first order in perturbation theory.
There will be various perimeter contributions such as those
of Fig.\ \ref{fig:Wilson1}a.
As we shall see, these are UV divergent and should
in principle be regulated.%
\footnote{
  As mentioned before, the dynamics of Theory 3 is UV finite.
  But the definition of the Wilson loop operator itself
  requires UV regularization.
}
But they don't depend on the separation $R$, and so
we can ignore them if we just focus on the $R$-dependence of the
Wilson loop expectation.
We will similarly ignore contributions that don't depend on
$\ttot$, such as Fig.\ \ref{fig:Wilson1}b.
$R$ dependence is generated by propagators which connect
different edges, such as in Fig.\ \ref{fig:Wilson1}c.
If we pick a reasonable gauge for doing perturbation theory,
then, in the large time ($\ttot$) limit for the Wilson loop,
we can neglect diagrams such as Fig.\ \ref{fig:Wilson1}d which attach to the
far-past or far-future ends of the loop.
In this case, the large-time Wilson loop, to lowest order,
is determined just by Fig.\ \ref{fig:Wilson1}c.

The primary example of an {\it un\/}reasonable gauge is $A_0=0$ gauge,
which is the gauge in which B\"odeker's effective theory (Theory 3) was
originally formulated.
There is no Fig.\ \ref{fig:Wilson1}c at all in $A_0=0$ gauge,
because the time-going Wilson lines only couple to $A_0$.
But $A_0=0$ gauge is a sick gauge for perturbation theory
in the first place.%
\footnote{
   That's because the free action in $A_0=0$ gauge
   has infinitely many zero modes
   associated with time-independent
   gauge transformations --- zero modes which are not properly treated
   in a perturbative expansion and which manifest as spurious
   non-integrable singularities in propagators.
}
We will instead work in Coulomb gauge.
The calculation of this section
is repeated in Appendix \ref{app:flow}\ in more general ``flow'' gauges,
which interpolate smoothly between Coulomb gauge and $A_0=0$ gauge.

The perturbative expansion of the action (\ref{eq:Lcoulomb})
describing B\"odeker's effective theory in Coulomb gauge is
\begin {equation}
   S_{\rm Coloumb} = \int dt\>d^3x \> {1\over 4 T} \left[
          \sigma \Bigl|\grad A_0\Bigr|^2
          + {1\over\sigma} \,
              \Bigl|\left(\sigma \partial_t - \grad^2\right) \A\Bigr|^2
          + O(\A^3)
          + \hbox{(ghosts)}
   \right] .
\end {equation}
One may read off the $A_0$ propagator (which is instantaneous in time):
\begin {equation}
   A_0^a \wiggle A_0^b
   ~~=~~ {2 T \over \sigma k^2} \> \delta^{ab} .
\end {equation}
Fig.\ \ref{fig:Wilson1}c then gives a contribution
to the Wilson loop of
\begin {eqnarray}
   d_\R^{-1} \, \delta \Wil
   &=& g^2 \, {\tr(T^a T^b) \over \tr(1)} \int_0^{\ttot} dt\>dt'\>
        \Bigdlangle A_0^a(t,0) A_0^b(t',\vecR) \Bigdrangle
\nonumber\\
   &=& - g^2 \, \CA \, \ttot \int_\k {2 T \over \sigma k^2} \,
   e^{i\k\cdot\vecR}
   = - {2 \alpha \CA T \over \sigma R} \, \ttot \,,	
\label {eq:Wilson1}
\end {eqnarray}
where $d_\R \equiv \tr(1)$ is the dimension of the representation
associated with the Wilson loop.
Because the $A_0$ propagator is instantaneous in time,
there are no crossed graphs at higher order and
this contribution exponentiates in the usual way to give
\begin {equation}
   \ln \Wil = - {2 \alpha \CA T \over \sigma R} \> \ttot	
           + \hbox{(higher order)}
\label{eq:lnW}
\end {equation}
in the large-time limit, up to terms independent of $R$ or $\ttot$.
In perturbation theory, at least, one sees that a large time-like Wilson loop
provides a gauge-invariant quantity from which one may extract $\sigma$.

One can automatically remove the perimeter terms independent of either
$R$ or $\ttot$ 
by taking ratios of Wilson loops:
\begin {equation}
   \ln\left[\Wil(R_1,t_{1}) \, \Wil(R_2,t_{2}) \over
            \Wil(R_1,t_{2}) \, \Wil(R_2,t_{1}) \right]
   = - {2\alpha \CA T \over \sigma} \left({1\over R_1} - {1\over R_2}\right)
               (t_{1}-t_{2})
     + \hbox{(higher order)}
   .						
\end {equation}
This ratio is free of UV divergences in Theory 3.

A warning is in order concerning the physical interpretation of the
result (\ref{eq:lnW}).
The result should be trustworthy whenever perturbation theory is reliable,
which means whenever $R \ll 1/g^2 T$.
Some readers may automatically associate Wilson loops with
the behavior
$\exp[-V(R)\,\ttot]$, with $V(R)$ interpreted as
the potential energy (or free energy) associated with two static
({\em i.e.}, infinitely massive) test charges separated by distance $R$.
However, this interpretation only applies to Wilson loops (or Polyakov lines)
in {\it Euclidean}\/ time and does not apply to the case at hand of
real-time loops at finite temperature.
In particular, (\ref{eq:lnW}) should not be interpreted as
a $1/R$ potential between static test charges for $R \ll 1/g^2T$,
which would be inconsistent with Debye screening for $R \gg m^{-1} = O(1/g T)$.
In fact, we know no simple physical interpretation of real-time
Wilson loops at finite temperature.

Though we have only discussed a perturbative analysis of Wilson loops,
that is good enough for matching Theory 2 to Theory 3,
whose physics only differs at scales $k \gtrsim \gamma = O(g^2 T \ln)$,
where the physics is still perturbative (by a logarithm).
Here and henceforth, in order of magnitude estimates
the abbreviation ``$\ln$'' is shorthand for $\ln(1/g)$.
As discussed in our companion paper \cite{overview}, and in earlier works by
Braaten and Nieto \cite{Braaten&Nieto}, matching may be performed by
formally computing the same quantity {\it perturbatively}\/ in both theories,
in the presence of some common infrared regulator.
We shall consider Wilson loops with $R \gg 1/\gamma$, so that they
are firmly in the region of validity of both theories, and we shall
regulate the infrared behavior using dimensional regularization.
For simplicity, we'll focus on the formal limit
$R \to \infty$ in the context of our IR-regulated perturbative calculation.
(The order of limits is important: the $\ttot \to \infty$ limit is to be
taken first, so as not to invalidate the previous discussion.)


\subsubsection{Relation to $\sigma$ in Theory 2}

As far as the $A_0$ 
propagator is concerned,
the perturbative expansion of the action
(\ref{eq:S2}) is much like that of Theory 3 except that the
color conductivity $\sigma$ becomes momentum dependent.
Specifically, the Coulomb gauge propagator for $A_0$ is now
\begin {equation}
   A_0^a \wiggle A_0^b
   ~~=~~ {2 T \over k^2 \, \sigmaL(k)} \, \delta^{ab} ,
\label{eq:A0prop0}
\end {equation}
where
\begin {equation}
   \sigmaL(k) \equiv \hat k_i \, \sigo_{ij}(\k) \, \hat k_j \,,
\end {equation}
and $\sigo_{ij} \equiv \sig_{ij}\bigl|_{\A=0}$
is given by (\ref{eq:sigmaD}) and (\ref{eq:Gdef})
with the covariant derivative $\D$ replaced by $i\k$:
\begin {eqnarray}
   \sigo_{ij}(\k)
   &=& m^2 \lim_{\Lambda\to\infty} \Bigl\langle v_i
             [ \v \cdot i\k + \deltaC + \Lambda \hat P_0 ]^{-1}
             v_j \Bigr\rangle
\nonumber\\
   &=&  m^2 \left[
      \langle v_i \hat G_0(\k) v_j \rangle
      - \langle v_i \hat G_0(\k) \rangle \langle \hat G_0(\k) \rangle^{-1}
        \langle \hat G_0(\k) v_j \rangle
   \right] ,
\label{eq:sigmak}
\end {eqnarray}
Here,
\begin {equation}
   \hat G_0(\k) \equiv [\v\cdot\grad + \deltaC]^{-1} 
            = [\v\cdot i\k + \deltaC]^{-1} .
\label {eq:G0def}
\end {equation}

The first-order contribution to the Wilson loop, analogous to
(\ref{eq:Wilson1}), is then
\begin {equation}
   d_\R^{-1} \, \delta \Wil
   = - g^2 \CA \ttot \int_\k {2 T \over k^2 \, \sigmaL(k)} \, e^{i\k\cdot\vecR}
   .
\label{eq:dWilson1}
\end {equation}
One should expect that the large $R$ behavior is dominated by the
small $k$ behavior of the integrand, and this is indeed the case,
up to corrections suppressed by powers of $g$.
(See appendix \ref{app:Wilson} for an explicit argument.)
The result is then
\begin {equation}
   d_\R^{-1} \, \delta \Wil
   \to - {2 \alpha \CA T \over \sigmaL(0) R} \, \ttot	
\label{eq:Wilson1b}
\end {equation}
for $R \to \infty$.
More specifically, this limit is $R \gg \gamma^{-1}$, since $\deltaC$
provides
the only scale in the definition of $\sigo(\k)$ and the scale of
$\deltaC$ is $\gamma$.

From (\ref{eq:sigmak}), we have
\begin {equation}
   \sigo_{ij}(0)
   = m^2 \lim_{\Lambda\to\infty} \Bigl\langle v_i \,
             [ \deltaC + \Lambda \hat P_0 ]^{-1}
             v_j \Bigr\rangle
\end {equation}
As noted in section I,
$v_j \rangle$ has $l=1$ --- that is, it is a superposition
of $|1m\rangle$'s.
Recalling that $\deltaC$ is diagonal in the space of $|lm\rangle$'s,
as is $\hat P_0 = |00\rangle\langle00|$, we obtain
\begin {equation}
   \sigo_{ij}(0) = {m^2\over d \gamma_1} \, \delta^{ij}
   , \qquad \hbox {and}\qquad
   \sigmaL(0) = {m^2 \over d \gamma_1}
   \,,
\label{eq:sigo0}
\end {equation}
in $d$ spatial dimensions,
where $\gamma_1 \equiv \langle 1 | \deltaC | 1 \rangle$
is the $l=1$ eigenvalue of $\deltaC$.
[We've left the spatial dimension $d$ arbitrary in
(\ref{eq:sigo0}) because the generalization away from $d{=}3$
will be needed later when we dimensionally regularize.]
Comparison of the Wilson loop (\ref{eq:Wilson1b}) in Theory 2 and
(\ref{eq:Wilson1}) in Theory 3 then gives the leading-order result
for matching the two theories:
\begin {equation}
   \sigma
          \approx {m^2 \over d \gamma_1}
          \approx {m^2 \over d \gamma} ,
\label{eq:sigmaLO}
\end {equation}
where the last leading-log equality uses (\ref{eq:C11leading}).
This is precisely the leading-log result for Theory 3 originally derived by
B\"odeker.
This result could have been very quickly derived from the starting
point of (\ref{eq:deltaCleading}) without all this
discussion of Wilson loops.
This approach based on Wilson loops does, however, provide
a conceptually clear framework for discussing sub-leading corrections.


\subsubsection{Next order in the loop expansion}

As mentioned earlier, perturbation theory at a scale $k$ is controlled
by the loop expansion parameter $g^2 T/k$.
A matching calculation between Theory 2
and Theory 3 is a calculation of physics at the interface $k \sim \gamma$
below which Theory 3 is valid, and so the loop expansion for matching
calculations will be an expansion in $g^2 T/\gamma \sim [\ln(1/g)]^{-1}$.
To go to next-to-leading-log order in the determination of $\sigma$, we
must therefore go to the next order in the loop expansion for the
Wilson loops.

A nice simplification occurs in Coulomb gauge.  Consider the quadratic
pieces of the $\E \, \sig(\D) \E$ term in the action (\ref{eq:S2})
for Theory 2:
\begin {equation}
   \E \sig(\D) \E
   = - A_0 \, \grad \cdot \sig(\grad) \grad A_0
     - 2 A_0 \, \grad \cdot \sig(\grad) \, \partial_t \A
     - \A \, \sig(\grad) \, \partial_t^2 \A
     + O(\A^3) .
\label {eq:EEexp}
\end {equation}
Rotation invariance implies that $\nabla_i [\sig(\grad)]_{ij}$
must be proportional to $\nabla_j$
(since $[\sig(\grad)]_{ij}$ can only involve
terms proportional to either $\delta_{ij}$ or $\nabla_i \nabla_j$).
Therefore, 
the second term in (\ref {eq:EEexp}) which connects
$A_0$ and $\A$ 
actually involves $\grad\cdot\A$ and hence vanishes
in Coulomb gauge.
Consequently, with this gauge choice the propagator does not mix
$A_0$ and $\A$.

\begin {figure}[t]
\vbox{
   \begin {center}
       \epsfig{file=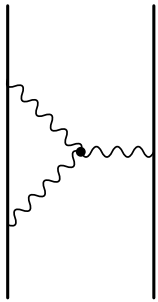,scale=0.85}
   \end {center}
   \caption{
       A next-order correction to Fig.\ \ref{fig:Wilson1}c that does not
       appear in Coulomb gauge.
       \label{fig:threeA0}
   }
}
\end {figure}

Now notice that the entire action (\ref{eq:S2}) is quadratic
in $A_0$ --- there are no $A_0^3$ or higher terms.
Taken together, this means that in Coulomb gauge there is no diagram
such as Fig.~\ref{fig:threeA0} contributing to the Wilson loop.
Instead, the next-to-leading-order diagrams all have the form of
self-energy corrections to the propagator in our leading-order diagram
Fig.\ \ref{fig:Wilson1}c.

\begin {figure}[t]
\vbox{
   \begin {center}
      \epsfig{file=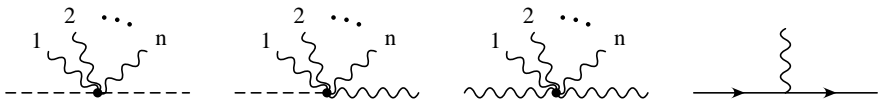,scale=0.8}
   \end {center}
   \caption{
       The Coulomb-gauge interaction vertices of action (\ref{eq:S2})
       of Theory 2.
       Dashed lines represent $A_0$, wavy lines $\A$,
       and solid lines the gauge-fixing ghost $\eta$.
       $n$ is any positive integer.
       \label{fig:vertices}
   }
}
\end {figure}

\begin {figure}
\vbox{
   \begin {center}
       \epsfig{file=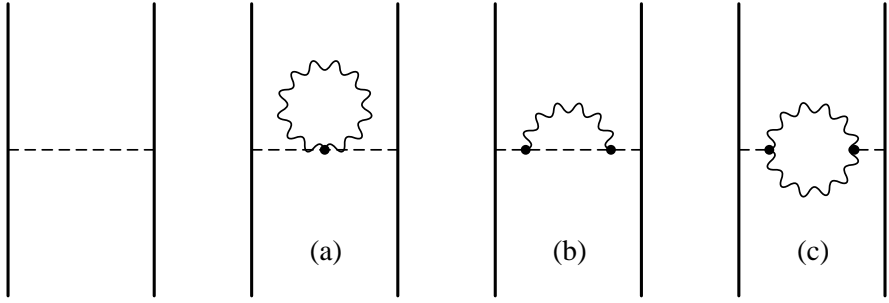,scale=0.85}
   \end {center}
   \caption{
       The first-order and next-order contributions to the Wilson
       loop in Coulomb gauge.
       \label{fig:diags}
   }
}
\end {figure}

We shall henceforth visually distinguish $A_0$ and $\A$ propagators
when drawing
Feynman diagrams, representing $A_0$ propagators as dashed lines and
$\A$ propagators as wavy lines.
An expansion of the Theory 2 action (\ref{eq:S2}) in powers
of $\A$ will give interaction vertices of the forms shown in
Fig.\ \ref{fig:vertices}.
In Coulomb gauge, the one-loop corrections to the Wilson loop are
then given by Fig.\ \ref{fig:diags}.
(We have left out diagrams involving
tadpoles, as these vanish by CP symmetry.)
This diagrammatic result applies to Theory 3 as well
as to Theory 2.

In Coulomb gauge, the calculation of the Wilson loop at this order
now reduces to the evaluation of the one-loop self-energy
$\Pi_{00}(\omega,\k)$ of $A_0$.
In fact, for the large Wilson loops discussed earlier, all that will
be important is the $\omega{=}0$, $\k \to 0$ behavior of the self-energy.
Absorbing $\Pi_{00}$ into the $A_0$ propagator (\ref{eq:A0prop0})
and repeating the
argument that led to (\ref{eq:Wilson1b}), we
have
\begin {equation}
   \sigmaL(k) \to \sig_{\rm L}^{(\eff)}(k)
   = \sigmaL(k) + {2T\over k^2} \, \Pi_{00}(0,k) \,,
\end {equation}
and
\begin {equation}
   d_\R^{-1} \, \delta \Wil
   \to
   - {2\alpha \CA T \over \sig_{\rm L}^{(\eff)}(0) R} \> \ttot \,, 
\end {equation}
in Theory 2.  Corresponding expressions hold for Theory 3,
with $\sigmaL(k)$ replaced by
$\sigma$.
Matching the two theories then gives
\begin {equation}
   \sigma = {m^2\over d\gamma_1}
      + \lim_{k\to0} \left\{
        \left[{2T\over k^2} \, \Pi_{00}(0,k)\right]_{\mbox{\small (Theory 2)}}
      - \left[{2T\over k^2} \, \Pi_{00}(0,k)\right]_{\mbox{\small (Theory 3)}}
        \right\}
      + O\left(\sigma \over \ln^2\right) .
\label{eq:match23a}
\end {equation}
Here and henceforth, all results are in Coulomb gauge unless explicitly
stated otherwise.

It must be emphasized that, because this is a matching calculation,
the $k\to0$ limits of the individual $\Pi_{00}/k^2$ terms
above are to be understood as
taken in the presence of an infrared regulator.
Suppose, for the sake of discussion, that $\Pi_{00}(k)/k^2$ were
given by
\begin {equation}
    k^{-2} \int_\p {k^2 \over p^2 \, |\k+\p|^2} \, ,
\label {eq:LimExample}
\end {equation}
where $\p$ was the spatial part of some loop momentum.
The integral is perfectly finite and so does not appear to require
any IR regularization.
Without IR regularization, the result of (\ref{eq:LimExample}) goes
like $k^{-1}$ by dimensional analysis, and the $k\to0$ limit is not
well defined.
Now imagine a simple momentum cut-off $M$ on small loop momenta.
The small $\k$ limit of (\ref{eq:LimExample}) then behaves as
$M^{-1}$ rather than $k^{-1}$, and there is no problem with the
limit.  In general, in the presence of an IR regulator, we can
formally expand integrands in $\k$, so that, for example,
\begin {equation}
    \lim_{\k\to0}k^{-2} \int_\p {k^2 \over p^2 \, |\k+\p|^2}
    = \int_\p {1\over p^2} \,,
\label {eq:LimExample2}
\end {equation}
in any IR regularization scheme.
In dimensional regularization the
result of this particular example is especially simple:
(\ref{eq:LimExample2}) is zero.


\subsection{Calculating \boldmath$\Pi_{00}(0,k{\to}0)$ in Theory 2}

\subsubsection{Perturbative expansion of the action}

We will now be explicit about the perturbative expansion of the
action (\ref{eq:S2}) for Theory 2.  Since we are working in
Coulomb gauge, there is an additional ghost piece of
$\bar\eta \grad\cdot\D \eta$ in the action but this is
irrelevant since no ghosts enter any of the diagrams in Fig.~\ref{fig:diags}
that we need to calculate.
In fact, we will be parsimonious in our discussion of the expansion
and only explicitly keep track of those terms which are relevant to
the specific diagrams of Fig.\ \ref{fig:diags},
namely the quadratic terms plus the $A_0 \A \A$, $A_0 A_0 \A$, and
$A_0 A_0 \A \A$ interactions.

Before expanding the action (\ref{eq:S2}), it is helpful to first make
some simplifications.  In particular, write $\E = \D A_0 - \dot \A$
and expand $\E \sig \E$ as
\begin {equation}
   \int d^d x \> \E \sig \E
     = \int d^d x \left[ - A_0 \, \D\sig\D \, A_0 + 2 A_0 \, \D\sig \, \dot\A
          + \dot\A \sig \dot\A \right] .
\label {eq:EsE0}
\end {equation}
From the expression (\ref{eq:sigmaD}) for $\sig$, we have
\begin {equation}
   \D \sig =
   m^2 \left(
      \langle \v\cdot\D \, \hat G \v\rangle
      - \langle \v\cdot\D \, \hat G \rangle \langle \hat G \rangle^{-1} 
                \langle \hat G \v \rangle
   \right) .
\label{eq:EsE1}
\end {equation}
Now comes a trick we will use repeatedly.
Since $\deltaC$ annihilates $\v$-independent states, as discussed in the
sec.\ \ref{sec:deltaC}, we have $\langle \deltaC = 0$ and can rewrite
\begin {equation}
   \langle \v\cdot\D
   = \langle (\v\cdot\D + \deltaC)
   = \langle \hat G^{-1} .
\label{eq:trick0}
\end {equation}
So
\begin {eqnarray}
   \langle \v\cdot\D \, \hat G &=& {}\langle\,, \qquad \hbox{and}\qquad
   \hat G \, \v\cdot\D \rangle = {}\rangle .
\end {eqnarray}
Eq.\ (\ref{eq:EsE1}) then simplifies to
\begin {equation}
   \D \sig =
   - m^2 \langle \hat G \rangle^{-1} 
                \langle \hat G \v \rangle
   ,
\end {equation}
and
\begin {equation}
   \D \sig \D =
   - m^2 \langle \hat G \rangle^{-1} 
   .
\end {equation}
We may now rewrite (\ref{eq:EsE0}) as
\begin {equation}
   \int d^d x \> \E \sig \E
     = \int d^d x \left[ m^2 A_0 \langle \hat G \rangle^{-1} A_0
          - 2 m^2 A_0 \langle \hat G \rangle^{-1}
                         \langle \hat G \v\rangle \cdot \dot\A
          + \dot\A \sig \dot\A \right] .
\label{eq:EsE2}
\end {equation}
The perturbative expansion is now generated by formally expanding the
expression (\ref{eq:Gdef}) for $G$:
\begin {equation}
   \hat G \equiv [\v\cdot\D + \deltaC]^{-1}
   = \hat G_0 - g \hat G_0 \v\cdot\A \hat G_0
          + g^2 \hat G_0 \v\cdot\A \hat G_0 \v\cdot\A \hat G_0
          - \cdots ,
\end {equation}
with $G_0$ given by (\ref{eq:G0def}).
(We will not explicitly show the factors of $\mu^{\eps/2}$ that accompany
$g$ in dimensional regularization.)
Then
\begin {eqnarray}
   \langle G \rangle^{-1} =
   \vevGo^{-1}
   &+& g \vevGo^{-1} \langle \Go \vA \Go \rangle \vevGo^{-1}
\nonumber\\ 
   &+& g^2 \vevGo^{-1} \langle \Go \vA \Go \rangle \vevGo^{-1}
                 \langle \Go \vA \Go \rangle \vevGo^{-1}
\nonumber\\ 
   &-& g^2 \vevGo^{-1} \langle \Go \vA \Go \vA \Go \rangle \vevGo^{-1}
\nonumber\\
   &+& O(g^3 \A^3) \,.
\label{eq:vevGinv}
\end {eqnarray}

Now focus for a moment on the expansion of the
$A_0 \langle \hat G \rangle^{-1} \langle\hat G\v\rangle \cdot \dot\A$
term in (\ref{eq:EsE2}), and consider in particular the leading-order
contribution to $\langle\hat G\v\rangle\cdot \dot\A$.
By rotation invariance, $\langle\hat G_0 v_i\rangle$ must have a factor of
$\nabla_i$, since $\grad$ is the only vector quantity appearing in
$G_0$.  Combining this with
\begin {equation}
  \langle\hat G_0 \v\rangle \cdot\grad =
  \langle\hat G_0 (\v\cdot\grad + \deltaC) \rangle = 1
\end {equation}
then yields%
\footnote{
  This formula does not easily generalize to $\langle G \v \rangle$ because
  of the non-commutativity of the covariant derivatives contained in $G$.
}
\begin {equation}
  \langle\hat G_0 \v\rangle = \langle \v G_0 \rangle = {\grad \over \nabla^2}
  \, .
\label{eq:vGo}
\end {equation}
Since we are in Coulomb gauge, we thus have
$\langle\hat G_0\v\rangle \cdot\dot\A = 0$.
So the leading term of the expansion vanishes,
and therefore
\begin {equation}
  A_0 \langle \hat G \rangle^{-1} \langle\hat G\v\rangle \cdot \dot\A
  = - g A_0 \vevGo^{-1} \langle \Go \vA \Go \v\rangle \cdot \dot\A
    + O(g^2 A_0 \A^3) .
\end {equation}

Putting everything together with the other terms in the action
(\ref{eq:S2}), and keeping track only of the terms that are needed for
the diagrams of Fig.\ \ref{fig:diags},
yields $S = S_{\rm free} + S_{\rm int}$, with
\begin {mathletters}
\begin {eqnarray}
   S_{\rm free} &=&
       \int\! dt \, d^d x \, {1\over 4T} \biggl\{
          m^2 A_0 \vevGo^{-1} A_0
          + \dot\A \, \sig(\grad) \, \dot\A
          + \A \grad^2 [\sig(\grad)]^{-1} \grad^2 \A
          + \mbox{\small (ghosts)}
       \biggr\} ,
\label{eq:S2xfree}
\\
   S_{\rm int\phantom{e}} &=&
       \int\! dt \, d^d x \, {1\over 4T} \biggl\{
          m^2 A_0 \left(\langle G\rangle^{-1} - \vevGo^{-1}\right) A_0
          + 2 g m^2 A_0 \vevGo^{-1} \langle \Go \vA \Go \v\rangle \cdot \dot\A
\nonumber\\ && \hspace{25.5em} {}
          + \mbox{\small (not needed)}
       \biggr\} ,
\label{eq:Sint}
\end {eqnarray}
\end {mathletters}%
where $\langle G \rangle^{-1}$ is expanded as shown in (\ref{eq:vevGinv}).
The perturbative expansion of the $L_1[\A]$ term of the action
(\ref{eq:S2}) falls into the ``not needed'' category since the
diagrams of Fig.\ \ref{fig:diags} do not contain any interaction vertices
involving only $\A$ and not $A_0$,
and any correction to the $\A$ propagator
induced by $L_1[\A]$ will be of sub-leading order
in logarithms.


\subsubsection{Propagators}

The perturbative $\sig(\grad)$ appearing in $S_{\rm free}$ can
be simplified a bit.
In momentum space, we earlier called it $\sigo(\k)$, given by
(\ref{eq:sigmak}):
\begin {equation}
   \sigo_{ij}
   =  m^2 \Bigl(
      \langle v_i \Go v_j \rangle
      - \langle v_i \Go \rangle \vevGo^{-1} \langle \Go v_j \rangle
   \Bigr) ,
\label{eq:sigmak2}
\end {equation}
Now note that the first term is transverse, because
\begin {equation}
   \nabla_i \langle v_i \Go v_j \rangle
   = \langle(\grad\cdot\v + \deltaC) \, \Go v_j \rangle
   = \langle v_j \rangle
   = 0
\end {equation}
(and similarly $\langle v_i \Go v_j\rangle \nabla_j = 0$).
So we may rewrite
\begin {equation}
   m^2 \langle v_i \, \Go(\k) \, v_j \rangle = \sigmaT(k) \, \PT^{ij}(\hat\k) ,
\label{eq:sigmaT}
\end {equation}
where we introduce (perturbative) transverse and longitudinal projection
operators
\begin {eqnarray}
   \PT^{ij}(\k) &\equiv& \delta^{ij} - \hat k^i \hat k^j ,
\\
   \PL^{ij}(\k) &\equiv& \hat k^i \hat k^j .
\end {eqnarray}
$\sigmaT$ may then be expressed as
\begin {equation}
   \sigmaT(k) = m^2 \, {\langle v_i \, \Go(\k) v_i \rangle \over d-1} \,.
\end {equation}
Similarly, the second term in (\ref{eq:sigmak2}) is purely longitudinal,
by (\ref{eq:vGo}).  So
\begin {equation}
   \sigo(\k) = \sigmaT(k) \, \PT(\hat\k) + \sigmaL(k) \, \PL(\hat\k) ,
\end {equation}
with
\begin {equation}
   \sigmaL(k) = {m^2\over k^2} \, \langle \Go(\k) \rangle^{-1} .
\label{eq:sLG0}
\end {equation}
From
(\ref{eq:sigo0}) for the low-momentum limit $\sigo(0)$ we find
\begin {equation}
   \sigmaT(0) = \sigmaL(0) = {m^2\over d \gamma_1} \,.
\label{eq:sig0}
\end {equation}

Because we are in Coulomb gauge, the longitudinal sector does not
contribute to $\dot\A \, \sig(\grad) \, \dot\A$ or
$\A [\sig(\grad)]^{-1} \A$.  So we can replace $\sig$ by $\sigmaT$
in the free action (\ref{eq:S2xfree}).
The resulting propagator for $\A$ is
\begin {eqnarray}
   A_i^a \wiggle A_j^b
   ~~=~~ S_{ij}(\omega,\k) \, \delta^{ab}
   ~&\equiv&~ {2 T \, \sigmaT(k) \over
               \left|i\omega\,\sigmaT(k) + k^2\right|^2}
         \,  \PT^{ij}(\hat\k) \, \delta^{ab}
\nonumber\\
   &=&~ {2 T \, \sigmaT(k) \over
               [\omega\,\sigmaT(k)]^2 + k^4}
         \,  \PT^{ij}(\hat\k) \, \delta^{ab} .
\label{eq:Aprop}
\end {eqnarray}
The relations between these propagators and the retarded,
equilibrium, or other types of propagators are discussed in Appendix
\ref{app:props}.
The propagator for $A_0$ is the same as (\ref{eq:A0prop0})
(although we are now representing this propagator by a dashed line),
\begin {equation}
   A_0^a \dashes A_0^b
   ~~=~~ {2T\over k^2 \, \sigmaL(k)} \, \delta^{ab}
   ~=~ {2 T \over m^2} \, \langle\Go(\k)\rangle \, \delta^{ab} . \kern .8in
\label{eq:A0prop}
\end {equation}


\subsubsection {Transposition}

It will be convenient to be able to rewrite interaction terms
(\ref{eq:Sint}) as their transposes.
Under transposition in $\x$/color space,
$\D^\trans = -\D$, and so
\begin {equation}
   \langle\hat G\rangle^\trans
                 = \Bigl\langle [\v\cdot\D + \deltaC]^{-1} \Bigr\rangle^\trans
                 = \Bigl\langle [-\v\cdot\D + \deltaC]^{-1} \Bigr\rangle
                 = \langle\hat G\rangle ,
\end {equation}
where the last equality follows from taking $\v \to -\v$ in the the
$\v$-average $\langle\cdots\rangle$.
The terms on the right-hand side of the perturbative expansion of
(\ref{eq:vevGinv}) must also be symmetric, and this can be explicitly
verified by recalling that the $\A$'s there are really
color matrices $\matA  = \A^a T^a$, which are anti-symmetric
because $T^b_{ac} = f^{abc}$.  So, for instance,
\begin {equation}
   \langle \Go \v\cdot\matA \Go \rangle^\trans
      = - \langle \Go^\trans \v\cdot\matA \Go^\trans \rangle
      = \langle \Go \v\cdot\matA \Go \rangle
\end {equation}
where the last equality again follows by $\v$-parity and where
we have implicitly used the fact that $\v$ and $\deltaC$ are
symmetric in $\v$-space.

Now let us check the transposition of the
$A_0 \vevGo^{-1} \langle\Go\v\cdot\matA\Go\v\rangle \cdot\dot\A$
interaction in (\ref{eq:Sint}).
Being a little more explicit about color indices than previously,
and placing an under-tilde on the $\A$ which is
to be interpreted as a color matrix, the interaction is
\begin {equation}
  A_0^a \, \langle \Go^{ab} \rangle^{-1}
  \langle\Go^{bc} \, \v\cdot\matA^{cd} \, \Go^{de} \, \v\rangle
  \cdot\dot\A^e .
\end {equation}
Its transpose is
\begin {equation}
  - \dot\A \cdot
  \langle\v\Go^\trans\v\cdot\matA\Go^\trans\rangle
  \langle \Go^\trans \rangle^{-1} A_0
  = - \dot\A \cdot \langle\v\Go\v\cdot\matA\Go\rangle \vevGo A_0 .
\end {equation}
In summary, this interaction term can be written in either of two ways:
\begin {equation}
  A_0 \vevGo^{-1} \langle\Go\v\cdot\matA\Go\v\rangle \cdot\dot\A
  = - \dot\A \cdot \langle\v\Go\v\cdot\matA\Go\rangle \vevGo A_0 .
\end {equation}


\subsubsection{Analysis of diagrams}


Instead of presenting general Feynman rules for all the various
vertices in the effective theory, and applying these rules to
the diagrams of Fig.~\ref{fig:diags}, we will
simply write the expressions for
the loop diagrams directly by treating $S_{\rm int}$ as a perturbation
in the path integral, expanding the exponential
$\exp(-S_{\rm int})$, and
explicitly taking all possible Wick contractions of the fields.
For the case at hand, this is far more convenient.
The loops (a-c) in Fig.\ \ref{fig:diags} represent the expressions
\begin {mathletters}
\begin {eqnarray}
   - \Pi_{00}^{\rm(a)} &=&
   2 \!\left(\!-{g^2m^2\over 4T}\right) \!
   \vevGo^{-1} \Bigl(
       \langle\Go\v\cdot
           \underwick{\A\Go\rangle \vevGo^{-1} \langle\Go\v\cdot\A}
           \Go \rangle
       - \langle\Go\v\cdot
           \underwick{\A\Go\v\cdot\A}
           \Go\rangle
   \Bigr) \vevGo^{-1} \!,
\nonumber\\
\label{eq:Pia}
\\
   - \Pi_{00}^{\rm(b)} &=&
   4 \!\left(\!- {gm^2\over 4T}\right)^2
   \Bigl[ \vevGo^{-1} \langle \Go \v\cdot
   \underwick{
      \A \Go \rangle \vevGo^{-1}
      \overwick{A_0 \Bigr] \otimes \Bigl[A_0}
      \vevGo^{-1} \langle \Go \v\cdot\A
   }
   \Go\rangle \vevGo^{-1} \Bigr] ,
\nonumber\\
\label{eq:Pib}
\\
   - \Pi_{00}^{\rm(c)} &=&
   \left(\!- {gm^2\over 2T}\right)^2
   \vevGo^{-1} \Bigl[
      \langle\Go\v\cdot
   \underwick{
      \A \Go \v\cdot
      \overwick{ \dot\A \rangle \otimes \langle - \v\cdot \dot\A }
      \Go \v\cdot\A
   }
   \Go \rangle
\nonumber\\ && \hspace{7em}
   {} + \langle \Go \v\cdot
   \underoverwick
     { \A \Go \v\cdot }
     { \dot\A \rangle \otimes \langle - \v\cdot\dot\A \vphantom{\Go}}
     { \Go \v\cdot\A }
   \Go \rangle \Bigr] \vevGo^{-1} .
\label {eq:Pic}
\end {eqnarray}
\end {mathletters}%
The $\otimes$ notation above denotes where strings of color index contractions
end.  Specifically, $(\cdots) \otimes (\cdots)$ represents the color matrix
$(\cdots)^a (\cdots)^b$, where $a$ and $b$ are adjoint color indices.%
\footnote{
   So, for example,
   $\langle\Go \vA \Go \v\cdot\dot\A\rangle
    \otimes
    \langle \v\cdot\dot\A \Go \vA \Go \rangle$
   represents
   $\langle\Go^{ac} \vA^{cd} \Go^{de} \v\cdot\dot\A^e\rangle
    \langle \v\cdot\dot\A^f \Go^{fg} \vA^{gh} \Go^{hb} \rangle$.
}

Using the $A_0$ propagator (\ref{eq:A0prop}), one can rewrite
(\ref{eq:Pib}) as
\begin {equation}
   - \Pi_{00}^{\rm(b)} =
   - 2 \left(- {g^2m^2\over 4T}\right)
   \Bigl[ \vevGo^{-1} \langle \Go \v\cdot
   \underwick{
      \A \Go \rangle \vevGo^{-1} \langle \Go \v\cdot\A
   }
   \Go\rangle \vevGo^{-1} \Bigr] .
\end {equation}
It neatly cancels the first term in (\ref{eq:Pia}) to leave
\begin {equation}
   - \Pi_{00}^{\rm(a+b)} \equiv
   - \left[\Pi_{00}^{\rm(a)} + \Pi_{00}^{\rm(b)}\right] =
   {g^2m^2\over 2T} \,
   \vevGo^{-1}
   \langle\Go\v\cdot
           \underwick{\A\Go\v\cdot\A}
           \Go\rangle
   \vevGo^{-1} .
\end {equation}
The next step is to use the $\A$ propagator (\ref{eq:Aprop}) and put
everything into frequency and momentum space:%
\footnote{
   The relative minus sign in (\ref{eq:moosepoop}) arises from the
   time derivatives in (\ref{eq:Pic}), which give $(-i p_0)(-i p_0)$
   for the first term, and $(-i p_0)(+i p_0)$ for the second.
}
\begin {eqnarray}
   \left[\Pi_{00}^{\rm(a+b)}(0,k)\right]_{ab} &=&
     - {g^2m^2\over 2T} \int_{p_0,\p}
     \langle\GGok v_i T^c_{ad}\,\Gop\,v_j T^c_{db}\,\GGok\rangle
     \, S_{ij}(-p_0,\k{-}\p) \,,
\\
   \left[\Pi_{00}^{\rm(c)}(0,k)\right]_{ab} &=&
     {g^2m^4\over 4 T^2} \int_{p_0,\p}
     \langle \GGok v_i T^c_{ad} \Gop v_j \rangle \,
     p_0^2 \, S_{i\bar\imath}(-p_0,\k{-}\p)
           \, S_{j\bar\jmath}(p_0,\p)
\nonumber\\ && \kern 1.8cm {} \times
     \left[
       \langle v_{\bar\jmath}\,\Gop\,v_{\bar\imath} T^c_{db}\,\GGok \rangle
       - \langle v_{\bar\imath} \,\Go(\k{-}\p)\, v_{\bar\jmath}
              T^d_{cb}\, \GGok \rangle 
     \right] ,
\label{eq:moosepoop}
\end {eqnarray}
where we find it convenient to introduce
\begin {equation}
   \GGok \equiv {\Gok \over \langle \Gok \rangle} \,.
\end {equation}
Remember that we are interested only in the case of zero external
frequency $k_0$.
The loop frequency integrals are easy to do using the explicit
form (\ref{eq:Aprop}) of $S_{ij}$, and give
\begin {equation}
   \int_{p_0} S_{i j}(-p_0,\q)
   = {T  \, \PT^{ij}(\hat\q)\over q^2} \, ,
\end {equation}
\begin {equation}
   \int_{p_0} p_0^2 \, S_{i k}(p_0,\q_1) \, S_{j l}(-p_0,\q_2)
   = {2 T^2 \, \PT^{ik}(\hat\q_1) \, \PT^{jl}(\hat\q_2)
         \over q_1^2 \,\sigmaT(q_2) + q_2^2 \,\sigmaT(q_1)} \, .
\end {equation}
Simplifying the color factors, we then have
\begin {eqnarray}
   \Pi_{00}^{\rm(a+b)}(0,k) &=&
     {\CA g^2m^2\over 2} \int_\p
     \langle\GGok\,v_i\,\Gop\,v_j\,\GGok\rangle \,
     {\PT^{ij}(\widehat{\k{-}\p}) \over |\k{-}\p|^2} \,,
\label{eq:Piab1}
\\
   \Pi_{00}^{\rm(c)}(0,k) &=&
     - {\CA g^2m^4\over 2} \int_{\p}
     \langle \GGok\,v_i\,\Gop\,v_j \rangle \,
     {\PT^{i\bar\imath}(\widehat{\k{-}\p}) \, \PT^{j\bar\jmath}(\hat\p)
         \over |\k{-}\p|^2 \,\sigmaT(p) + p^2 \,\sigmaT(|\k{-}\p|)}
\nonumber\\ && \kern 2.2cm {} \times
     \left[
       \langle v_{\bar\jmath}\,\Gop\,v_{\bar\imath}\,\GGok \rangle
       + \langle v_{\bar\imath}\,\Go(\k{-}\p)\,v_{\bar\jmath}\,\GGok \rangle 
     \right] .
\label{eq:Pic1}
\end {eqnarray}


\subsubsection{The $\k\to0$ limit}

We now want to extract the $k\to0$ limit of $\Pi_{00}(k)$ through
$O(k^2)$,
as this is required for
the Wilson loop matching formula (\ref{eq:match23a}).
This section is somewhat tedious, and some readers may wish to skip to
the result (\ref{eq:klim1}).

To classify the various terms in the $\k$ expansion,
it will help to rewrite the total self-energy given by
(\ref{eq:Piab1}) and (\ref{eq:Pic1}) as
\begin {equation}
   \Pi_{00}(0,k) =
   {\CA g^2 \over 2} \langle \GGok \, \Ok \GGok \rangle ,
\label{eq:Odef0}
\end {equation}
with
\begin {eqnarray}
   \Ok &=&
     m^2 \int_\p \biggl\{
        v_i\,\Gop\,v_j \,
       {P^{ij}_{\k{-}\p} \over |\k{-}\p|^2}
\nonumber\\ && \hspace{2em}
     - m^2 v_i\,\Gop\,v_j \rangle \,
       {P^{i\bar\imath}_{\k-\p} P^{j\bar\jmath}_\p
         \over |\k{-}\p|^2 \sigma_\p + p^2 \sigma_{\k-\p}}
     \left[
       \langle v_{\bar\jmath}\,\Gop\,v_{\bar\imath}
       + \langle v_{\bar\imath}\,\Go(\k{-}\p)\,v_{\bar\jmath} 
     \right]
   \biggr\} ,
\label{eq:Odef}
\end {eqnarray}
where in this section we use the abbreviations
\begin {equation}
   \sigma_\p \equiv \sigmaT(p),
   \qquad\qquad
   P^{ij}_\p \equiv \PT^{ij}(\hat\p),
\end {equation}
to keep formulas more compact.
%
The operator
$\O$ is symmetric in $\v$-space.  This is manifest for all
the terms except the one involving $\Go(\k{-}\p)$, and it is easily
seen for that term by making the change of integration variable
$\p \to \k{-}\p$.

It is useful to start with the small $\k$ expansion of
$\GGo(\k)\rangle$.  
This expansion is derived in appendix \ref{app:G0x} and is
\begin {equation}
   \GGo(\k)\rangle = {\Go(\k)\rangle \over \langle \Go(\k) \rangle}
   = \left(1 - {\v\cdot i\k\over\gamma_1}\right)\Bigr\rangle + O(k^2) .
\label {eq:GG0x}
\end {equation}
We now turn to the operator $\O(\k)$, and
first consider $\O(\k)\rangle$.
From the definitions (\ref{eq:Odef}) of $\O(\k)$ and
(\ref{eq:sigmaT}) of $\sigma_\p = \sigmaT(p)$, we have
\begin {equation}
   \Ok\rangle =
     m^2 \int_\p v_i\,\Gop\,v_j \rangle \left\{
       {P^{ij}_{\k-\p} \over |\k{-}\p|^2}
       - {P^{ik}_{\k-\p} P^{jk}_\p \,
         (\sigma_\p + \sigma_{\k-\p})
         \over |\k{-}\p|^2 \sigma_\p + p^2 \sigma_{\k-\p}}
   \right\},
\label{eq:O1}
\end {equation}
and in particular
\begin {equation}
   \O({\bf 0}) \rangle = 0 .
\end {equation}
This means that $\Ok\rangle$ is $O(k)$, and it is the
reason that we only need the expansion of
$\GGo (\k)$ thru $O(k)$ and not $O(k^2)$.
Specifically, using the symmetry of $\O(\k)$ and $\GGo(\k)$ under
$\v$-space transposition, we can organize the small $k$ expansion of
(\ref{eq:Odef0}) as
\begin {equation}
   \Pi_{00}(k) = {\CA g^2\over 2} \left[
      \langle \Ok \rangle
      - 2 \gamma_1^{-1} \langle \v\cdot i\k \, \Ok \rangle
      + \gamma_1^{-2} \langle \v\cdot i\k \,\O({\bf 0})\,
             \v\cdot i\k \rangle
      + O(k^3)
   \right] .
\label{eq:reorg}
\end {equation}


\paragraph*{First term:}
Let's begin with $\langle \Ok\rangle$.
Starting with (\ref{eq:O1}), one can again use the definition
(\ref{eq:sigmaT}) of $\sigma_\p = \sigmaT(p)$ to find
\begin {eqnarray}
   \langle\Ok\rangle &=&
     \int_\p P^{ij}_\p P^{ij}_{\k-\p} \sigma_\p \left[
         {1 \over |\k{-}\p|^2}
       - {(\sigma_\p + \sigma_{\k-\p})
          \over |\k{-}\p|^2 \sigma_\p + p^2 \sigma_{\k-\p}}
     \right]
\nonumber\\
   &=&
     \int_\p P^{ij}_\p P^{ij}_{\k-\p} \,
         { \left(p^2 - |\k{-}\p|^2\right) \over |\k{-}\p|^2 } \,
         { \sigma_\p \sigma_{\k-\p}
           \over |\k{-}\p|^2 \sigma_\p + p^2 \sigma_{\k-\p} }
     \,.
\end {eqnarray}
Now symmetrize the integrand with respect to the change of integration
variable $\p \to \k{-}\p$:
\begin {eqnarray}
   \langle\Ok\rangle &=&
     \int_\p P^{ij}_\p P^{ij}_{\k-\p} \,
         { \left(p^2 - |\k{-}\p|^2\right)^2 \over 2p^2|\k{-}\p|^2 } \,
         { \sigma_\p \sigma_{\k-\p}
           \over \left(|\k{-}\p|^2 \sigma_\p + p^2 \,\sigma_{\k-\p}\right) }
\nonumber\\
   &=&
     (d{-}1) \int_\p
         { \left(\k\cdot\p\right)^2 \sigma_\p \over p^6 }
     + O(k^3)
\nonumber\\
   &=&
     {(d{-}1) \over d} \, k^2 \int_\p
         { \sigma_\p \over p^4 }
     + O(k^3) \,.
\label{eq:term1}
\end {eqnarray}


\paragraph*{Second term:}
For the
$\gamma_1^{-1} \langle \v\cdot i\k \, \Ok\rangle$ term
of (\ref{eq:reorg}), we need the
$O(k)$ piece of $\Ok\rangle$.
Its extraction can be simplified by rewriting (\ref{eq:O1}) as
\begin {eqnarray}
   \Ok\rangle &=&
     m^2 \int_\p v_i\,\Gop\,v_j \rangle \>\Biggl\{
         {1 \over |\k{-}\p|^2}
           \left[ P^{ij}_{\k-\p}
             - P^{ik}_{\k-\p} P^{jk}_\p \right]
\nonumber\\ && \hspace{8em}
       - { \left(|\k{-}\p|^2 - p^2\right) \sigma_{\k-\p}
           \over
           |\k{-}\p|^2
           \left(|\k{-}\p|^2 \,\sigma_\p + p^2 \,\sigma_{\k-\p}\right)}
         \, P^{ik}_{\k-\p} P^{jk}_\p
   \Biggr\}
\nonumber\\
   &=&
     m^2 \int_\p v_i\,\Gop\,v_j \rangle \> \Biggl\{
         {1 \over p^2}
           \left[ P^{ij}_{\k-\p}
             - P^{ik}_{\k-\p} P^{jk}_\p \right]
       + { \p\cdot\k \over p^4 } P^{ik}_\p P^{jk}_\p
   \Biggr\} + O(k^2)
\nonumber\\
   &=&
     m^2 \int_\p v_i\,\Gop\,v_j \rangle \,
         {1 \over p^4} \left[
           k^i p^j - 2\hat p^i \hat p^j \p\cdot\k + \delta^{ij} \p\cdot\k
         \right]
     + O(k^2) \,.
\end {eqnarray}
Therefore,
\begin {equation}
   \langle \v\cdot i\k \, \Ok\rangle =
     m^2 \int_\p \langle v_l v_i\,\Gop\,v_j \rangle \,
         {i k^l \over p^4} \left[
           k^i p^j - 2\hat p^i \hat p^j \p\cdot\k + \delta^{ij} \p\cdot\k
         \right]
     + O(k^3) \,.
\label {eq:O3}
\end {equation}
Because of rotation invariance,
$\langle \v\cdot i\k \, \Ok\rangle$
depends only on the magnitude $k$ of $\k$,
so nothing is harmed by averaging
the integrand in (\ref{eq:O3}) over the direction $\hat\k$, giving
\begin {equation}
   \langle \v\cdot i\k \, \Ok\rangle =
     {m^2 k^2\over d} \!\int_\p {1\over p^4}\Bigl[
         \langle \Gop\,\v\cdot i\p \rangle
         - 2 \langle \v\cdot\hat\p \, \v\cdot\hat\p
                    \, \Gop \, \v\cdot i\p \rangle
    + \langle \v\cdot i\p \, v_i \,\Gop\,v_i \rangle
     \Bigr] .
\label {eq:O3b}
\end {equation}
The factors of $\Go(\p)$ can be eliminated from the
first and second terms by using the
trick (\ref{eq:trick0}), which in the present context is
\begin {equation}
  \Gop \, \v\cdot i\p\rangle =
  \Gop \, (\v\cdot i\p + \deltaC)\rangle = {}\rangle .
\end {equation}
Using the relations (\ref{eq:dC1}),
a similar manipulation shows that
\begin {equation}
   \langle \v\cdot i\p \, v_i \,\Gop
   = \langle v_i \v\cdot i\p \,\Gop
   = \langle v_i (\v\cdot i\p+\deltaC-\gamma_1) \,\Gop
   = \langle v_i [1 - \gamma_1 \, \Gop] \,,
\label{eq:trick3}
\end {equation}
and so the final term of (\ref {eq:O3b}) becomes
\begin {equation}
   \langle \v\cdot i\p \, v_i \,\Gop\, v_i \rangle
   = 1 - (d{-}1) \, {\gamma_1\over m^2} \, \sigma_\p .
\end {equation}
Putting everything together,
\begin {equation}
   \gamma_1^{-1} \langle \v\cdot i\k \, \Ok\rangle =
     {(d{-}1)\over d} \, k^2 \int_\p {1\over p^4}
         \left(2\sigma_0 - \sigma_\p\right) ,
\label{eq:term2}
\end {equation}
where $\sigma_0 = m^2/d\gamma_1$ is the value of $\sigma_\p$ at
$\p=0$, taken from (\ref{eq:sig0}).


\paragraph*{Third term:}
Finally, we pursue the
$\gamma_1^{-2} \langle \v\cdot i\k \, \O({\bf 0}) \, \v\cdot i\k \rangle$
term in the expansion (\ref{eq:reorg}).  We are again free to average
over the direction $\hat\k$, giving
\begin {equation}
   \langle \v\cdot i\k \, \O({\bf 0}) \, \v\cdot i\k \rangle =
   - {k^2\over d} \, \langle v_k \, \O({\bf 0}) \, v_k \rangle
   .
\label{eq:ksimp3}
\end {equation}
Now use the definition (\ref{eq:Odef}) of $\O(\k)$ to find:
\begin {eqnarray}
   \langle v_k \, \O({\bf 0}) \, v_k \rangle &=&
     \int_\p {m^2\over p^2} \biggl\{
        \langle v_k \, v_i\,\Gop\,v_j \, v_k \rangle
        P^{ij}_{\p}
\nonumber\\ && \hspace{1em}
     - m^2 \langle v_k \, v_i\,\Gop\,v_j \rangle \,
     {P^{i\bar\imath}_\p P^{j\bar\jmath}_\p
         \over 2 \sigma_\p}
     \left[
       \langle v_{\bar\jmath}\,\Gop\,v_{\bar\imath} \, v_k\rangle
       + \langle v_{\bar\imath}\,\Go({-}\p)\,v_{\bar\jmath}
                    \, v_k\rangle
     \right]
   \biggr\} .
\label{eq:O4}
\end {eqnarray}
The second term vanishes by the following symmetry argument.
First, use $\v\to-\v$ and the definition (\ref{eq:G0def}) of $\Go(\k)$
to rewrite
\begin {equation}
       \langle v_{\bar\jmath}\,\Gop\,v_{\bar\imath} \, v_k\rangle
       + \langle v_{\bar\imath}\,\Go({-}\p)\,v_{\bar\jmath}
                    \, v_k\rangle
       =
       \langle v_{\bar\jmath}\,\Gop\,v_{\bar\imath} \, v_k\rangle
       - (\bar\imath \leftrightarrow \bar\jmath) .
\label{eq:ijsym}
\end {equation}
By rotation invariance,
$\langle v_{\bar\jmath}\,\Gop\,v_{\bar\imath} \, v_k\rangle$
can only depend on the direction as
$\hat\p_{\bar\imath} \hat\p_{\bar\jmath} \hat\p_k$,
$\hat\p_{\bar\imath} \delta_{\bar\jmath k}$,
$\hat\p_{\bar\jmath} \delta_{\bar\imath k}$,
and
$\hat\p_{k} \delta_{\bar\imath \bar\jmath}$.
Every one of these possibilities either has a $\hat\p_{\bar\imath}$ or
a $\hat\p_{\bar\jmath}$, which will annihilate against the
transverse projections $P^{i\bar\imath}_\p P^{j\bar\jmath}_\p$
in (\ref{eq:O4}), or
else has a $\delta^{\bar\imath \bar\jmath}$, which vanishes by the
anti-symmetry of (\ref{eq:ijsym}) in $\bar\imath\bar\jmath$.
In summary, (\ref{eq:ksimp3}) and (\ref{eq:O4}) become simply
\begin {eqnarray}
   \langle \v\cdot i\k \, \O({\bf 0}) \, \v\cdot i\k \rangle &=&
     - {k^2\over d} \int_\p {m^2\over p^2} \,
        \langle v_k \, v_i\,\Gop\,v_j \, v_k \rangle
        P^{ij}_{\p} .
\label {eq:term3}
\end {eqnarray}
Simplifying this expression will require us to be a little more
systematic about the manipulations we have been using and is the
subject of a later section.  We end this one by combining the results
(\ref{eq:term1}), (\ref{eq:term2}), and (\ref{eq:term3}) for the
individual terms appearing in (\ref{eq:reorg}):
\begin {equation}
   \Pi_{00}(k) = - {\CA g^2 k^2 \over 2} \int_\p \left[
         {(d{-}1) \over d \, p^4} \, (4\sigma_0 - 3 \sigma_\p)
         + {\sigma_0\over \gamma_1 p^2} \,
            \langle v_k \, v_i\,\Gop\,v_j \, v_k \rangle
            P^{ij}_{\p}
   \right] + O(k^3) .
\label {eq:klim1}
\end {equation}


\subsubsection{Extracting UV and IR divergences}
\label{sec:IR}

As we shall see, the integral (\ref{eq:klim1}) giving $\Pi_{00}$ is
both infrared and ultraviolet divergent in three spatial dimensions.
We are using dimensional regularization, but it
will simplify our discussion of what to do with
$\langle v_k \, v_i\,\Gop\,v_j \, v_k \rangle$ if we can instead work
directly in three dimensions.
Therefore, we will now isolate the pieces of the integrand responsible
for the IR and UV divergences, so that we can subtract them and
evaluate the remainder as a finite integral in $d{-}3$.
Specifically, we will rewrite (\ref{eq:klim1}) as
\begin {equation}
   \Pi_{00}(k) = - {\CA g^2 k^2 \over 2} \left[
     \int_\p f_{\rm reg}(\p)
   + \int_\p f_{\rm IR}(\p)
   + \int_\p f_{\rm UV}(\p)
   \right]
   + O(k^3) ,
\label {eq:klim2}
\end {equation}
where
\begin {equation}
   f_{\rm reg}(\p) =
         {(d{-}1) \over d \, p^4} \, (4\sigma_0 - 3 \sigma_\p)
         + {\sigma_0\over \gamma_1 p^2}  \,
            \langle v_k \, v_i\,\Gop\,v_j \, v_k \rangle
            P^{ij}_{\p}
         - f^{\rm(IR)}(\p)
         - f^{\rm(UV)}(\p) .
\end {equation}
$f_{\rm IR}$ and $f_{\rm UV}$ will be chosen to (a) make
the $f_{\rm reg}$ integral finite, and so evaluable directly in
three dimensions, and (b) make
the $f_{\rm IR}$ and $f_{\rm UV}$
integrals analytically tractable in dimensional
regularization.


\paragraph*{IR behavior:}
In appendix \ref{app:G0x}, we show that $\Go(\p)$ has the small
$\p$ expansion
\begin {equation}
   \Go = {d\over\gamma_1 p^2} \, (\gamma_1-\v\cdot i\p) \,
                       \Po \, (\gamma_1 - \v\cdot i\p) + O(p^0) .
\end {equation}
We can thus expand the $\langle \v \v \Go \v \v\rangle$ term in
(\ref{eq:klim1}) as
\begin {eqnarray}
   {\sigma_0 \over \gamma_1 p^2} \langle v_k v_i \, \Gop \, v_j v_k \rangle
       P_\p^{ij}
   &=& {d \sigma_0 \over \gamma_1^2 p^4} \,
     \Bigl\langle v_k v_i \, (\gamma_1-\v\cdot i\p) \Bigr\rangle
     \Bigl\langle (\gamma_1- i\v\cdot\p) \, v_j v_k \Bigr\rangle \, P_\p^{ij}
         + O(p^{-2})
\nonumber\\
   &=& {(d{-}1)\over d} \, {\sigma_0 \over p^4} + O(p^{-2}) .
\end {eqnarray}
As discussed earlier, the $\p \to 0$ limit of $\sigma_\p$ is
$\sigma_0 = m^2/(d\gamma_1)$.  In Appendix \ref{app:G0x}, we show that
the small $\p$ corrections to $\sigma(p)$ are $O(p^2)$.
Putting everything together, we may then choose
\begin {equation}
   f_{\rm IR}(\p) =
      2 \, {(d{-}1) \over d} \, {\sigma_0 \over  p^4} .
\end {equation}
In dimensional regularization, the integral of $f_{\rm IR}$ vanishes:
\begin {equation}
   \int_\p f_{\rm IR}(\p) = 0 .
\end {equation}


\paragraph*{UV behavior:}
For $\p \to \infty$, we can treat $\deltaC$ as a perturbation to
$\v\cdot i\p$, giving $\Go \to (\v\cdot i\p)^{-1}$, except that we
will need a prescription for integrating over the pole $\v\cdot\p = 0$
in angular integrals.  The prescription is obtained by recalling
from section \ref{sec:deltaC} that $\deltaC$ is a non-negative operator.
So
\begin {equation}
   \Gop \to {1\over \v\cdot i\p + \varepsilon}
        = {\rm P.P.} \, {1\over\v\cdot i\p} + \pi \delta(\v\cdot\p) \,,
\label{eq:GoUV}
\end {equation}
where $\varepsilon$ is a positive infinitesimal and ${\rm P.P.}$ denotes
principal part.  (Higher-order corrections to this formula are discussed
in Appendix \ref{app:largep}.)
This limit then gives
\begin {equation}
   \sigma_\p = {m^2\over (d{-}1)} \, \langle v_i \,\Gop\, v_i \rangle
      \to {m^2\over (d{-}1)} \, \pi \langle \delta(\v\cdot\p) \rangle
      = {m^2\over (d{-}1)} \, {S_{d-2}\,\pi\over S_{d-1} \,p} ,
\end {equation}
and
\begin {equation}
   \langle v_k v_i \,\Gop\, v_j v_k \rangle P_\p^{ij}
   \to \pi \langle v_i \,\delta(\v\cdot\p) v_j \rangle P_\p^{ij}
   = \pi \langle \delta(\v\cdot\p) \rangle
   = {S_{d-2} \,\pi \over S_{d-1} \,p} ,
\end {equation}
where
\begin {equation}
   S_{d-1} \equiv
   {2 \pi^{d/2} \over \Gamma(d/2)}
\label{eq:Sd}
\end {equation}
is the surface area of a ($d{-}1$)-sphere
({\em e.g.}, $S_1 = 2\pi$ and $S_2 = 4\pi$).
The UV piece of the integrand (\ref{eq:klim1}) is therefore
\begin {equation}
   f_{\rm UV}(\p) \to {S_{d-2} \, \pi\sigma_0 \over S_{d-1}\, \gamma_1} \,
                      {1\over p^3} \,,
\end {equation}
and comes only from the $\langle \v \v \Go \v \v\rangle$ term.
We don't want $f_{\rm UV}$ to mess up our IR subtraction, so we will
cut it off in the infrared by choosing
\begin {equation}
   f_{\rm UV}(\p) = {S_{d-2} \,\pi\sigma_0 \over S_{d-1}\,\gamma_1} \,
                    {1\over p \,(p^2+M^2)} \,,
\end {equation}
where $M$ is arbitrary.
In dimensional regularization, the integral of $f_{\rm UV}$ is
\begin {equation}
   \int_\p f_{\rm UV}(\p) = {\Gamma\left(\eps/2\right) \over
                                      2 (4\pi)^{1-\eps/2}} \,
                            {\sigma_0 \over \gamma_1}
                            \left(\mu\over M\right)^\eps
                 = {\sigma_0\over 4\pi\gamma_1} \left[
                      {1\over\eps} + \ln\left(\bar\mu\over M\right)
                   + O(\eps) \right] ,
\end {equation}
where $\bar\mu$ is the \MSbar\ scale defined by
\begin {equation}
   \bar\mu = \mu \sqrt{4\pi \over e^{\gammaE}} \,.
\end {equation}

Putting everything together, Eq.~(\ref{eq:klim2}) for $\Pi_{00}(k)$ becomes
\begin {mathletters}
\label {eq:klim3}
\begin {equation}
   \Pi_{00}(k) = - {\CA \alpha \sigma_0 \over 2\gamma_1} \, k^2 \left[
     {1\over\eps} + \ln\left(\bar\mu\over M\right) +
     {4\pi\gamma_1 \over \sigma_0} \int_\p f_{\rm reg}(\p,M)
   \right]
   + O(k^3)  + O(\eps),
\end {equation}
where we may now set $d{=}3$ in
\begin {equation}
   f_{\rm reg}(\p,M) =
         {1 \over p^4} \left({\textstyle{4\over3}}\sigma_0 - 2\sigma_\p\right)
         + {\sigma_0\over \gamma_1 p^2} \,
            \langle v_k \, v_i\,\Gop\,v_j \, v_k \rangle \,
            P^{ij}_{\p}
         - {\pi\sigma_0\over 2\gamma_1} \, {1\over p(p^2+M^2)} \,.
\label {eq:freg3}
\end {equation}
\end {mathletters}%
This formula for $\Pi_{00}$ does not depend on the choice of $M$.


\subsubsection{Reducing
  $\langle v_{i_1} \cdots v_{i_m} \, \Gop \, v_{j_1} \cdots v_{j_m} \rangle$}
\label{sec:reduce}

In the previous section, we encountered several $\v$-averages of $\Gop$
flanked by
various factors of $\v$.  Any finite combination of $\v$'s can be rewritten as
a superposition of spherical harmonics $|lm\rangle$'s, and so we
can recast the problem of simplifying general expressions of the form
$\langle v_{i_1} \cdots v_{i_m} \, \Gop \, v_{j_1} \cdots v_{j_m} \rangle$
to the simplification of
$\langle l'm' | \Gop | l m\rangle$.

Choose the $z$-axis in the direction of $\p$.  
Then
\begin {equation}
   \langle l'm' | \Gop | l m\rangle =
   \langle l'm' | (i p v_z + \deltaC )^{-1} | lm \rangle .
\end {equation}
Now recall that $\deltaC$ is diagonal in $l$ and $m$, and note that
$v_z$ can change $l$ but does not change the azimuthal quantum number
$m$.  So
\begin {equation}
   \langle l'm' | \Gop | l m\rangle =
   \langle l'm | (i p v_z + \deltaC )^{-1} | lm \rangle \, \delta_{m m'}.
\end {equation}
We can derive a recursion relation in $l$ by writing
\begin {eqnarray}
   \delta_{ll'}
   &=& \langle l'm | (ipv_z+\deltaC) \Go | lm \rangle
\nonumber\\
   &=&
       ip \, \langle l'm | v_z | l''m \rangle \,\langle l''m | \Go | lm \rangle
           + \langle l' | \deltaC | l' \rangle \,
             \langle l'm | \Go | l m \rangle \,,
\end {eqnarray}
with an implied sum over $l''$.
Since
$v_z$ can only change $l$ by $\pm1$, this gives
\begin {eqnarray}
   \delta_{l l'}
   &=&
   ip \langle l'm | v_z | (l'{+}1)m \rangle \langle (l'{+}1)m | \Go | lm \rangle
   +
   ip \langle l'm | v_z | (l'{-}1)m \rangle \langle (l'{-}1)m | \Go | lm \rangle
\nonumber\\&& {}
   + \langle l' | \deltaC | l' \rangle \langle l'm|\Go|lm \rangle \,.
\end {eqnarray}
This defines a recursion relation which allows one to rewrite matrix
elements with higher $l'$ in terms of those with lower $l'$.
This recursion will end when $l'$ becomes as low as it can be consistent
with $m$---that is, at $l'=|m|$.  A similar recursion can be constructed for
$l$, and by use of these recursions, all matrix elements
$\langle l'm|\Go|lm\rangle$ can be rewritten in terms of
$\langle |m|m\Big|\Go\Big| |m|m\rangle$.
In fact, the case $m<0$ is related to the case $m>0$ by
$\v$-parity:
\begin {equation}
   \langle |m|, m \Big| \Gop \Big| |m|, m \rangle
   = \langle |m|,{-}m \Big| \Go(-\p) \Big| |m|,{-}m \rangle
   = \langle |m|,{-}m \Big| \Gop \Big| |m|,{-}m \rangle \,,
\end {equation}
where the final equality follows from $\v$-parity and the
$[\v\cdot i\p + \deltaC ]^{-1}$ structure of $G_0(\p)$.

We now turn to the specific
problem of rewriting the $\langle v_k v_i \,\Gop\, v_j v_k \rangle$
term of our expression (\ref{eq:klim1})
for $\Pi_{00}$ in terms of the $\langle mm|\Go|mm\rangle$.
A product of two $\v$'s is a combination of $l{=}2$ and $l{=}0$,
so we will be able to rewrite the expectation in terms of
$\langle 22| \Go | 22\rangle$ and $\langle \Go \rangle$.
The advantage of this rewriting is that later analysis of how
to evaluate expectations involving $\Go$ will be simpler and more
natural for
$\langle mm | \Go |mm \rangle$ than for
$\langle v_k v_i \Go v_j v_k \rangle$ directly.

It is easiest to work backwards from the explicit form
for $Y_{2,2}$ which gives
\begin {equation}
   |22\rangle = - \sqrt{15\over 32\pi} (\hat v_x + i \hat v_y)^2 \rangle \,.
\end {equation}
From this, one may easily check that
\begin {equation}
   \langle 22 | \Go(\p) | 22 \rangle
   = {15\over 8} \, \langle v_k v_i \, \Go(\p) v_j v_l \rangle
            \left( 2 P_\p^{ij} P_\p^{kl} - P_\p^{ik} P_\p^{jl} \right) ,
\label {eq:G22}
\end {equation}
for any operator $\Go(\p)$, remembering that we have chosen $\hat\p$ to be the
$z$ direction for the purpose of defining $|lm\rangle$.%
\footnote{
   The natural generalization to $d$ dimensions is
   \[
     \langle 22 | \Go(\p) | 22 \rangle
     \equiv
         {d \,(d{+}2)\over (d^2{-}1)(d{-}2)} \,
         \langle v_k v_i \, \Go(\p) v_j v_l \rangle
            \left[(d{-}1)\, P_\p^{ij} P_\p^{kl} - P_\p^{ik} P_\p^{jl} \right],
   \]
   where the relative coefficient of the two $P_\p P_\p$ terms
   is chosen so that contraction with $\delta_{ik}$ or $\delta_{jl}$ gives
   zero (so as to exclude the $l=0$ combinations of $\v\v$),
   and the overall
   normalization has been chosen so that replacing $\Go(\p)$ by 1 gives
   $\langle 22 | 22 \rangle = 1$.
}
From (\ref{eq:G22}), we have
\begin {eqnarray}
   \langle v_k v_i \Go v_j v_k \rangle P_\p^{ij}
   &=& \langle v_k v_i \Go v_j v_l \rangle P_\p^{ij}
              (P_\p^{kl} + \hat\p^k \hat\p^l)
\nonumber\\
   &=& {4 \over 15} \, \langle 22|\Go|22 \rangle
     + \langle v_k v_i \Go v_j v_l \rangle
             \left( \half P_\p^{ik} P_\p^{jl} + P_\p^{ij} \hat\p^k \hat\p^l
                    \right) .
\end {eqnarray}
The second term on the right-hand side can be simplified by expanding
$P_\p^{ij} = \delta^{ij} - \hat\p^i \hat\p^j$, and by repeated use
of the relation (\ref{eq:trick3}). 
The result is
\begin {equation}
   \langle v_k v_i \Go v_j v_k \rangle P_\p^{ij}
   = {4 \over 15} \, \langle 22|\Go|22 \rangle
     + {1\over2} \, \langle G_0 \rangle
     - {\gamma_1\over p^2} \left(
          {1\over 6} + {2\over 3} \, {\sigma_\p\over\sigma_0} \right)
   .
\label{eq:G22simp}
\end {equation}

The analogous derivation for $\langle 11 | \Go | 11 \rangle$ gives%
\footnote{
   The natural generalization to $d$ dimensions is
   $
     \displaystyle
     \langle 11 | \Go | 11 \rangle
     \equiv {d\over d{-}1} \, \langle v_i \, \Go v_j \rangle \,
            P_\p^{ij}
     = d \, \sigma_\p / m^2.
   $
}
\begin {equation}
   \langle 11 | \Go | 11 \rangle
   = {3\over2} \, \langle v_i \Go v_j \rangle \, P_\p^{ij}
   = {3\over2} \, \langle v_i \Go v_i \rangle
   = {3 \sigma_\p \over m^2} \,.
\label {eq:Sig11}
\end {equation}

To obtain the conductivity at NLLO, it is adequate to use leading-log
approximations to the propagators 
in our one-loop calculation
of $\Pi_{00}$.  That is, we only need to evaluate the integrand
$f_{\rm reg}$ in (\ref{eq:klim3}) at leading-log order.
At this order, if one factors out
the scale $\gamma$ of $\deltaC$,
functions like $\langle 22 | \Go(\p) | 22 \rangle$
may be re-expressed as purely numerical
functions of the single dimensionless variable $p/\gamma$.
Specifically, at leading-log order,
\begin {equation}
   \left\langle |m|\,m \Big| \Go(\p) \Big| |m|\,m \right\rangle
   \approx \gamma^{-1} \, \Sigma_{|m|}(p/\gamma)
   ,
\label{eq:G0mm}
\end {equation}
where
\begin {equation}
   \Sigma_m(\rho) \equiv
   \biggl\langle mm \bigg| \, {1\over i \rho v_z + \deltac}
                   \bigg| mm \biggr\rangle ,
\label{eq:Sigdef}
\end {equation}
and $\deltac$ is the leading-log result (\ref{eq:deltaCleading})
for $\deltaC/\gamma$:
\begin {equation}
   \delta c(\v,\v') =
       \deltaS(\v-\v')
       - {4\over\pi} \> {(\v\cdot\v')^2 \over \sqrt{1-(\v\cdot\v')^2}} 
   .
\label{eq:deltac}
\end {equation}
We can now combine this with (\ref{eq:klim3}) and (\ref{eq:G22simp})
to obtain
\begin {equation}
   \Pi_{00}(k) = - {\CA \alpha \sigma_0 \over 2\gamma} \, k^2 \left[
     {1\over\eps} + \ln\left(\bar\mu\over M\right) +
     I\!\left({M\over\gamma}\right)
   \right]
   \times \left[ 1 + O\left(\ln^{-1}\right) \right]
   + O(k^3) + O(\eps),
\label {eq:T2result}
\end {equation}
with
\begin {equation}
   I(\nu) = {2\over\pi} \int_0^{\infty} d\rho \left[
      {1\over 2} \left( \Sigma_0(\rho) - {3\over\rho^2}\right)
      + {8\over3\rho^2} \Bigl(1 - \Sigma_1(\rho)\Bigr)
      + {4\over 15} \, \Sigma_2(\rho)
      - {\pi \rho \over 2 (\rho^2+\nu^2)}
   \right] .
\label {eq:I}
\end {equation}
We have used (\ref {eq:Sig11}) to rewrite $\sigma_\p$
(at the order under consideration) as
$\sigma_0 \, \Sigma_1(p/\gamma)$.
We have ceased to distinguish between $\gamma_1$ and $\gamma$ in $\Pi_{00}$
since we are ignoring corrections to $\Pi_{00}$ suppressed by additional
powers of inverse logs [see (\ref{eq:C11leading})].
The terms in (\ref{eq:I}) have been arranged so that each term is
individually IR safe.
We will discuss how to evaluate $I(\nu)$ numerically in section
\ref{sec:SigFormula}.


\subsection{Matching to Theory 3}
\label{sec:final23match}

   We've now got $\Pi_{00}$ in Theory 2, but we still need $\Pi_{00}$ in
Theory 3, so that we can use the matching condition (\ref{eq:match23a}) to
determine the parameter $\sigma$ of Theory 3 at NLLO.  Fortunately, the
one-loop calculation in Theory 3 is trivial in dimensional regularization:
$\lim_{k\to0}[k^{-2} \, \Pi_{00}(k)] = 0$.
The reason is simple dimensional analysis.
Rescale the variables of the path integral (\ref{eq:S3}) for Theory 3 to
$\bar t = \sigma^{-1} t$, $\bar\A = T^{-1/2}\A$, and
$\bar A_0 = \sigma T^{-1/2} A_0$.
Here we will for once be explicit about the factors of $\mu^\eps$.
The path action can then be rewritten as
\begin {equation}
   S = {1\over 4} \int d\bar t \> d^d x
          \left| -\bar\E + \bar\D\times\bar\B \right|^2 ,
\end {equation}
where $\bar D_\nu = \bar\partial_\nu + g \mu^{\eps/2} T^{1/2} \bar A_\nu$.
In this form, the parameters of the theory appear only in the combination
$g \mu^{\eps/2} T^{1/2}$.
At one loop, the self-energy $\bar\Pi_{00} \equiv \sigma^{-1} T \Pi_{00}$
of $\bar A_0$ must be proportional to
$g^2 \mu^\eps T$, which has mass dimension $1+\eps$.
But $\lim_{k\to0} [\k^{-2} \, \bar\Pi_{00}(k)]$
has mass dimension zero, and there are no other dimensionful parameters
in the problem that can make up the discrepancy in  mass dimension!
Consistency then forces
$\lim_{k\to0} [\k^{-2} \, \bar\Pi_{00}(k)]$ = 0 in dimensional regularization.
Such simplicity is the standard virtue of dimensional regularization for
matching calculations \cite{Braaten&Nieto,dimreg}.

   Our matching condition (\ref{eq:match23a}) and our Theory 2 result
(\ref{eq:T2result}) then give the color conductivity $\sigma$ at NLLO:
\begin {equation}
   \sigma = {m^2\over d\gamma_1}
       \left\{ 1
          - {\CA \alpha T \over \gamma} \left[
                 {1\over\eps} + \ln\left(\bar\mu\over\gamma\right) + I(1)
             \right]
          + O(\ln^{-2})
       \right\} ,
\label{eq:fsigma}
\end {equation}
where, for the sake of definiteness, we have fixed $M=\gamma$.
As one can see, the only information we will need about $\deltaC$ at
NLLO is the value of $\gamma_1$.

We now turn to methods for evaluating the functions $\Sigma_m(\rho)$ and
so evaluating the numerical constant $I(1)$ from (\ref{eq:I}).


\subsection{Evaluation of
  \boldmath$\Sigma_m(\rho) \approx \gamma \,\langle mm | \Gop | mm\rangle$}
\label{sec:SigFormula}

The dimensionless functions $\Sigma_m(\rho)$ were defined in (\ref{eq:Sigdef})
as $\langle mm | (i v_z \rho + \deltac)^{-1} | mm \rangle$, where
$\deltac$ is the leading-log result for $\deltaC/\gamma$.
As discussed in section \ref{sec:reduce},
the operators $v_z$ and $\deltaC$ both
preserve $m$, $\deltaC$ preserves $l$ as well, and $v_z$ changes $l$
by $\pm 1$.  For fixed $m$, the operator $i v_z \rho + \deltac$ may
therefore be considered as a tri-diagonal matrix in the $|lm\rangle$ basis
where $l = m$, $m{+}1$, $m{+}2$, ...\thinspace:
\begin {equation}
   i v_z \rho + \deltac = 
   \pmatrix{
      c_m       & i\bm_m\rho    &               &               &        \cr
\vphantom{\strut^{\strut}}
      i\bm_m\rho& c_{m+1}       & i\bm_{m+1}\rho&               &        \cr
\vphantom{\strut^{\strut}}
                & i\bm_{m+1}\rho& c_{m+2}       & i\bm_{m+2}\rho&        \cr
\vphantom{\strut^{\strut}}
                &               & i\bm_{m+2}\rho& c_{m+3}       & \ddots \cr
                &               &               & \ddots        & \ddots \cr
   } ,
\label {eq:SigMat}
\end {equation}
where
\begin {equation}
   c_l \equiv \langle l | \deltac | l \rangle ,
   \qquad
   \bm_l \equiv \langle lm | v_z | (l{+}1) m \rangle .
\end {equation}
$\Sigma_m(\rho)$ corresponds to the upper-left element of the inverse of
the matrix (\ref{eq:SigMat}).
Inverting tri-diagonal matrices is particularly simple, and there is a
continued-fraction formula for this element:%
\footnote
    {%
    This easily 
    follows from
    iterating the formula for the inverse of an $(N{+}1)\times (N{+}1)$
    matrix in terms of the inverse of its lower-right $N\times N$ block:
    $
	\left(
	\begin {array}{ll}
	    a & b^T \! \\ c & d
	\end {array}
	\right)^{-1}
	=
	\left(
	\begin {array}{cc}
	    \phantom- \alpha & -\alpha \, b^T d^{-1} \\
	    -\alpha d^{-1} c & (1 {+} \alpha d^{-1} c b^T) \, d^{-1}
	\end {array}
	\right),
    $
    with
    $\alpha \equiv (a - b^T d^{-1} c)^{-1}$.
    Here, $a$ is a scalar, $b$ and $c$ are $N$-component (column) vectors,
    and $d$ is an $N\times N$ matrix.
    }
\begin {equation}
   \Sigma_{m}(\rho) = 
   {1\over\displaystyle c_m +
      {\strut \bigl(\bm_m\rho\bigr)^2\over\displaystyle c_{m+1} +
         {\strut \bigl(\bm_{m+1}\rho\bigr)^2\over\displaystyle c_{m+2} +
            {\strut \bigl(\bm_{m+2}\rho\bigr)^2 \over\displaystyle \cdots}
   }}} \,.
\label{eq:cf}
\end {equation}
Note that $\Sigma_m(\rho)$ is even in $\rho$. 

All we need now are explicit formulas for the coefficients $c_l$ and $\bm_l$.
Eq.\ (\ref{eq:cf}) may then be used for numerical evaluation of
$\Sigma_m(\rho)$.%
\footnote
    {
    From Eqs.~(\ref {eq:bm}) and (\ref {eq:cl}), one may read off that
    $b_l^{(m)} \to \half$ and $c_l \to 1$ as $l\to\infty$.
    Consequently, the tail of the continued fraction,
    $
        X \equiv
        \lim_{k\to\infty} c_{m+k} + (b_{m+k}^{(m)} \rho)^2/(\cdots)
    $,
    satisfies
    $
        X = 1 + (\rho/2)^2 X^{-1}
    $,
    implying that
    $
        X = \half + \half \sqrt{1+\rho^2}
    $.
    The continued fraction (\ref {eq:cf}) may be evaluated quite accurately
    by replacing its tail at large but finite $l$ by this value.
    }
The $\bm_l$ are given by
\begin {equation}
   \bm_l = \sqrt{(l+1)^2-m^2 \over 4(l+1)^2 - 1} \,.
\label {eq:bm}
\end {equation}
The $c_l$ may be evaluated from the expression (\ref{eq:deltac})
for $\delta c(\v,\v')$ as
\begin {equation}
   c_l = \langle \delta c(\v,\v') \, P_l(\v\cdot\v') \rangle_{\v\v'}
       = 1 - {2\over\pi} \int_{-1}^{+1} dz \> {z^2 P_l(z)\over \sqrt{1-z^2}}
       \,,
\end {equation}
where $P_l(z)$ are Legendre polynomials.  The integral vanishes if $l$ is odd
and gives%
\footnote{
   We evaluated the integral using Eq.\ 2.17.2 of Ref.\ \cite{prudnikov2},
   and verified the result numerically.
}
\begin {eqnarray}
   c_{2n} &=& 1 - 2
       \left[ {(2n)! \over 2^{2n} \, (n!)^2} \right]^2
       \left(1 + {1\over (2n)^2 + 2n - 2}\right) ,
\nonumber\\
   c_{2n+1} &=& 1 \,.
\label {eq:cl}
\end {eqnarray}
The $c_l$ are all non-negative, as was claimed in section
\ref{sec:deltaC}.  Note that $c_0$ vanishes, as it must.

The procedure for numerical evaluation of $\Sigma_m(\rho)$ is to
compute the continued-fraction formula (\ref{eq:cf}) with some
upper cut-off $l_{\rm max}$ on $l$, and then repeat the calculation,
doubling $l_{\rm max}$ each time until the answer converges.
This procedure becomes inefficient for very large $\rho$, however, because
it then requires rather large $l_{\rm max}$ for good convergence.
For very large $\rho$, it is more convenient to use
asymptotic formula for $\Sigma_m(\rho)$, which are derived and presented
in Appendix \ref{app:largep}.

The final result of numerical evaluation of the integral (\ref{eq:I}) that
defines $I(\nu)$, using numerical evaluation of the functions
$\Sigma_m(\rho)$ as described above, is
\begin {equation}
   I(1) = 2.8380\cdots .
\label{eq:Inumber}
\end {equation}



\section {Matching theory 1 to theory 2}
\label {sec:sigma12}

Our next task is to determine the operator $\deltaC$ of Theory 2,
appearing in (\ref{eq:2a}).  Specifically, we want
$\gamma_1 = \langle 1 | \deltaC | 1 \rangle$ to leading order in $g$
(and all orders in logs).
We will follow the general matching strategy used previously.
We will temporarily introduce an infrared cut-off,
then compute the total effective collision operator $\delta C_{\rm tot}$
in both theories, formally expanded to leading order in perturbation theory,
and then determine what bare collision operator $\deltaC$ appearing in
Eq.~(\ref{eq:2a}) of Theory 2 is required for the results to match.
We will again use dimensional regularization to regulate the infrared.


\subsection {$\delta C_{\rm tot}$ in the underlying theory (Theory 1)}

There are now a variety of methods for computing the effective
collision operator
at leading order in the underlying short-distance theory
\cite{bodeker,bodekereps,Blog1,manuel,basagoiti}.%
\footnote{
   Another interesting analytic approach that gives $\gamma_1$ at
   leading log order is
   that of sections 2 and 3 of
   ref.\ \cite{moore}.  It is not clear to us how to extend this
   approach beyond leading log order, and in particular how to obtain the
   terms of $\deltaC(\v,\v')$ that are not proportional to
   $\delta(\v-\v')$.
}
Previous authors have only extracted the leading log
piece of their result because the leading-order result is formally
log divergent in the infrared.
Having regulated the infrared, we shall instead extract the entire thing.%
\footnote{
   Ref.\ \cite{bodekereps} also discusses doing this but does not go
   so far as to extract an explicit result.
}
So, one may now follow, in $d$ spatial dimensions, one's favorite
method of the references just cited.
The method we're most intimately familiar with is our own,
so our discussion will most closely parallel the presentation in
Ref.\ \cite{Blog1}.

\begin {figure}
\vbox{
   \begin {center}
       \epsfig{file=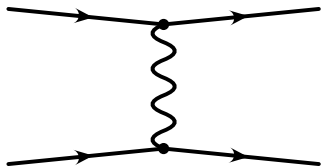,scale=0.50}
   \end {center}
   \caption{
       The dominant scattering process:
       $t$-channel gauge boson exchange.  The solid lines represent any
       sort of hard particles,
       including the non-Abelian gauge bosons themselves.
       \label{fig:tchannel}
   }
}
\end {figure}

At leading order in $g$, $\delta C$ is generated by $2 \leftrightarrow 2$
collisions of
hard particles, mediated by semi-hard ($q_0 \le q \lesssim m$)
t-channel gluon exchange, such as depicted in Fig.\ \ref{fig:tchannel}.
One finds%
\footnote{
   Here, as well as in Eqs.\ (\ref{eq:1}) and (\ref{eq:2}) defining
   either effective theory,
   $m$ should be understood as the $d$-dimensional Debye mass.
}
\begin {equation}
   \delta C \, W(\v) =
   {\CA m^2 T \over 2 g^2 \mu^\eps}
          \left\langle \int_\q |{\cal M}(\v,\v',\q)|^2 \>
          [W(\v) - W(\v')] \right\rangle_{\v'}
   ,
\label {eq:dc1}
\end {equation}
where ${\cal M}$ is the amplitude for a t-channel collision between
hard particles
with velocities $\v$ and $\v'$, mediated by a semi-hard gauge boson with
momentum $\q$.
(This interpretation of ${\cal M}$ reverts to a more fundamental picture
than that of Theory 1, interpreting $W$ as made up of individual hard
particles.  A derivation that is more directly in the framework of
Theory 1 may be found in Ref.\ \cite{bodeker}.)

If the integrand of (\ref {eq:dc1}) is separated into two pieces,
the coefficient of the first term, proportional to $W(\v)$,
is the same expression one obtains in a
leading order calculation of the hard gauge boson damping rate
\cite{gammag},
\begin {equation}
   \gamma = 
   {\CA m^2 T \over 2 g^2 \mu^\eps}
   \left\langle \int_\q |{\cal M}(\v,\v',\q)|^2
                 \right\rangle_{\v'}.
\label{eq:gama}
\end {equation}
The overall coefficient in front of $\int_\q |{\cal M}|^2$ in (\ref{eq:gama})
simply represents the results of group factors and the integration of
the magnitude $|\p'|$ in the calculation of $\gamma$ based on
Fig.\ \ref{fig:tchannel}.
The dependence on exactly what species of hard particles are present
is completely isolated in the value of the Debye screening mass.
The formula (\ref{eq:dc1}) for $\delta C \, W(\v)$ may equivalently
be converted to a formula for the kernel $\delta C(\v,\v')$ itself,
\begin {equation}
   \delta C(\v,\v') =
   {\CA m^2 T \over 2 g^2 \mu^\eps} \left[
          \left\langle \int_\q |{\cal M}(\v,\v',\q)|^2 \right\rangle_{\v'}
              \deltaS(\v{-}\v')
          - \int_\q |{\cal M}(\v,\v',\q)|^2
   \right]
   .
\label {eq:dc2}
\end {equation}

We have seen that, for the calculation of the NLLO conductivity, we will
not need the full form of $\delta C$, but will only need the matrix element
$\gamma_1 \equiv \langle 1 | \delta C | 1\rangle 
= \langle v^i \delta C v^i \rangle$.
We will begin with a general analysis of $\delta C$, however, and
specialize to $\gamma_1$ later.

If there were no screening effects to consider, the scattering amplitude
${\cal M}$ would be the classic Coulomb amplitude such that
\begin {equation}
   \int_\q |{\cal M}|^2
    = g^4 \mu^{2\eps} \int {d^{4-\eps} Q\over (2\pi)^{4-\eps}} \>
           {\left|v_\mu \, {\delta^{\mu\nu} \over Q^2} \, v'_\nu\right|^2}
           \> 2\pi \delta(Q\cdot v) \> 2\pi \delta(Q\cdot v')
    ,
   \qquad \hbox{(no screening)}
\end {equation}
where $Q = (q^0,\q)$ and $v = (1,\v)$.
The $\delta$-functions are simply
$q^0,q \ll T$ approximations of the constraints that the final-state
hard particles be on shell.
To account for screening of the exchanged semi-hard gluon, however, we must
replace this by
\begin {equation}
   \int_\q |{\cal M}|^2
    = g^4 \mu^{2\eps} \int {d^{4-\eps} Q\over (2\pi)^{4-\eps}} \>
      \left| v_\mu \left(
          {\PT^{\mu\nu}(Q)\over Q^2 + \PiT(Q)} +
          {\PL^{\mu\nu}(Q)\over Q^2 + \PiL(Q)}
      \right) v'_\nu \right|^2
      \,2\pi \delta(Q\cdot v) \> 2\pi \delta(Q\cdot v')
    \,,
\label{eq:Msqr1}
\end {equation}
where $\PT$ and $\PL$ are the transverse and longitudinal projection
operators,
\begin {eqnarray}
   \PT^{\mu\nu}(Q) &=& \cases{
          0, & $\mu=0$ or $\nu=0$; \cr
          {\displaystyle{\delta^{ij} - {q^i q^j\over q^2}}}, \qquad
                         & $\mu=i$ and $\nu=j$, \cr
   }
\\
   \PL^{\mu\nu}(Q) &=& 
        \left(g^{\mu\nu} - {Q^\mu Q^\nu\over Q^2}\right) - \PT^{\mu\nu} ,
\end {eqnarray}
and where $\PiT$ and $\PiL$ are the transverse and
longitudinal pieces of the standard, leading-order, hard thermal loop
self-energy \cite{pi}.
This self-energy depends on $Q$ only through the ratio $\lambda \equiv q^0/q$.

We will review the explicit formula for $\Pi(Q)$ below, but it's
worth first examining some qualitative features.
The longitudinal sector is screened for small $Q$ by Debye/plasmon effects.
The transverse sector, however, is unscreened in the $\lambda \to 0$ limit,
reflecting the fact that charged plasmas do not screen static magnetic
fields.  This lack of static screening is responsible for the logarithmic
infrared sensitivity that generates the usual leading-log result
for the color conductivity.  The logarithmic divergence appears only in
the purely transverse term of the squared amplitude (\ref{eq:Msqr1}).
It will be convenient to isolate that divergence by rewriting
(\ref{eq:Msqr1}) as
\begin {equation}
   \int_\q |{\cal M}|^2 =
   \int_\q \left(|{\cal M}|^2 - |{\cal M}|^2_\IR\right)
          + \int_\q |{\cal M}|^2_\IR \,,
\label {eq:split}
\end {equation}
where $|{\cal M}|^2_\IR$ is a small $q^0$ (small $\lambda$)
limiting form of $|{\cal M}|^2$
that we shall discuss in a moment.
Our strategy is to arrange that the integral of
$|{\cal M}|^2_\IR$ be simple enough
to evaluate in dimensional regularization, whereas the first term in
(\ref{eq:split}) will be completely finite and evaluable directly
in $d=3$ dimensions.

To continue,
switch integration variables from $q^0$ to $\lambda$ and, making use of
the $\delta$-functions, rewrite (\ref{eq:Msqr1}) as
\begin {eqnarray}
   \int_\q |{\cal M}|^2
    &=& 2\pi g^4  \mu^{2\eps} \int_{-1}^{+1} d\lambda \int_\q {1\over q}
      \left|
          {\v\cdot\v' - \lambda^2\over q^2(1-\lambda^2) + \PiT(\lambda)} +
          {-1 + \lambda^2 \over q^2(1-\lambda^2) + \PiL(\lambda)}
      \right|^2
\nonumber\\ && \hspace{15em} {} \times
      \delta(\lambda-\hat\q\cdot\v) \> \delta(\lambda-\hat\q\cdot\v')^{\strut}
      \,.
\label{eq:Msqr2}
\end {eqnarray}
The angular integration over $\hat\q$ is equivalent to replacing the
pair of delta functions by their angular average.  We will denote
this average, in $d$-dimensions, by
\begin {equation}
   f_d(\lambda,\v\cdot\v') \equiv
   \Bigl\langle \delta(\lambda-\hat\q\cdot\v) \>
           \delta(\lambda-\hat\q\cdot\v') \Bigr\rangle_{\hat\q} \,.
\end {equation}

We will implement our split (\ref{eq:split})
by extracting the small $\lambda$ behavior of
Eq.\ (\ref{eq:Msqr2}),
\begin {equation}
   \int_\q |{\cal M}|^2_\IR
    \equiv 2\pi g^4  \mu^{2\eps} \int_{-1}^{+1} d\lambda \int_\q {1\over q}
      \left|
          {\v\cdot\v'\over q^2 + \PiT^\IR(\lambda)}
      \right|^2
      \delta(\hat\q\cdot\v) \> \delta(\hat\q\cdot\v')
      ,
\label {eq:MsqrIR0}
\end {equation}
where $\PiT^\IR$ is the limiting small $\lambda$ behavior of $\PiT$,
to be discussed explicitly below.


\subsubsection{The IR piece}

In terms of
\begin {equation}
   f_d(0,\v\cdot\v') =
   \Bigl\langle
      \delta(\hat\q\cdot\v) \> \delta(\hat\q\cdot\v')
   \Bigr\rangle_{\hat\q}
   = {1-\eps\over 2\pi \sqrt{1-(\v\cdot\v')^2}}  \,,
\end {equation}
we have
\begin {equation}
   \int_\q |{\cal M}|^2_\IR
    = 2\pi (\v\cdot\v')^2 f_d(0,\v\cdot\v') \, g^4
      \mu^{2\eps} \int_{-1}^{+1} d\lambda \int_\q
          {1\over q \left| q^2 + \PiT^\IR(\lambda) \right|^2}
      \,.
\label{eq:MsqrIR1}
\end {equation}
We now need the form of $\PiT^\IR(\lambda)$.  The usual $d=3$ result for
the small frequency behavior of $\PiT$ is $-{i\pi\over4}\, m^2 \lambda$.
But we need the result in $3-\eps$ dimensions.  One could derive this
directly by evaluating a one-loop thermal diagram in the fundamental
quantum field theory, but let's instead derive it in the usual way from
effective theory 1.  Working at leading order in $g$, formally solving
the $W$ equation (\ref{eq:1a}),
and plugging into the Maxwell equation
(\ref{eq:1b}) gives
\begin {equation}
   \partial_\nu F^{\mu\nu}
   = m^2 \left\langle v^\mu v^i \over v\cdot\partial + \varepsilon
         \right\rangle_{\v} E^i
   ,
\end {equation}
where the $\varepsilon$ is simply an infinitesimal prescription specifying
retarded behavior.
Comparing with $\partial_\nu F^{\mu\nu} + \Pi^{\mu\nu} A_\nu = 0$,
the components of the self-energy can then be read off,
among which
\begin {equation}
   \Pi^{ij}(q) = -m^2 q^0 
                  \left\langle {v^i v^j \over -q^0 + \v\cdot\q -i\varepsilon} 
                  \right\rangle_{\v}.
\label {eq:pi1}
\end {equation}
We want the transverse self-energy
\begin {equation}
   \PiT = {1\over d{-}1} \> \tr\,(\PT \Pi) .
\label{eq:pi2}
\end {equation}
One can evaluate this for arbitrary frequency,%
\footnote{
   The result is
   $\PiT = {m^2\over d{-}1} \left[ \lambda^2 + (1{-}\lambda^2)
                  \> {}_2F_1({1\over2},1;{d\over2};\lambda^{-2}) \right]$
   and
   $\PiL = m^2 (1{-}\lambda^2) \left[1 - 
                  \, {}_2F_1({1\over2},1;{d\over2};\lambda^{-2}) \right]$.
}
but here we're only interested in the small frequency limit
$\PiT^\IR$.  Taking $q^0$ small, (\ref{eq:pi1}) and (\ref{eq:pi2}) yield
\begin {equation}
   \PiT^\IR =
        -{m^2 q^0\over d{-}1}
        \left\langle
            {1-(\v\cdot\hat\q)^2 \over \v\cdot\q - i\varepsilon}
        \right\rangle_\v
   =    
        -{m^2 q^0\over d{-}1}
        \left\langle i\pi\>\delta(\v\cdot\q) \right\rangle_\v
   =
        -{i\pi S_{d-2}\over (d{-}1) \, S_{d-1}} \, m^2 \lambda
   =
        -{i\pi\over4} \, \kappa \, m^2 \lambda ,
\end {equation}
where the area $S_{d-1}$ of a ($d{-}1$)-sphere is given by (\ref{eq:Sd}),
and where
\begin {equation}
   \kappa \equiv 
        {2\over\pi}\, B\left(\half,{\textstyle{d\over2}}\right)
        = 1 + \left[\ln 2 - \half\right] \eps + O(\eps^2) .
\end {equation}
We can now do the momentum integral in (\ref{eq:MsqrIR1}) using
\begin {equation}
   \int_\q {1\over q \, |q^2 + i c|^2} =
   \int_\q {1\over q \, (q^4 + c^2)} =
   {S_{d-1} \over (2\pi)^d} \, {\pi/4 \over \cos(\eps\pi/4)} \, |c|^{(d-5)/2} ,
\end {equation}
to give
\begin {eqnarray}
   \int_\q |{\cal M}|^2_\IR
   &=& 
       {S_{d-1} \over (2\pi)^{d-1}} \, {\pi/4 \over \cos(\eps\pi/4)} \, 
       (\v\cdot\v')^2 f_d(0,\v\cdot\v') \,
       g^4 \mu^{2\eps} \int_{-1}^{+1} d\lambda
       \left| {\pi\over4} \kappa m^2 \lambda \right|^{-1-(\eps/2)}
\nonumber\\
   &=& - 
       {\pi S_{d-1} \over \eps(2\pi)^{d-1} \cos(\eps\pi/4)} \,
       (\v\cdot\v')^2 f_d(0,\v\cdot\v')\,
       g^4 \mu^{2\eps}
       \left({\pi\over4}\kappa m^2 \right)^{-1-(\eps/2)}
\nonumber\\
   &=& 
       {2g^4\mu^\eps (\v\cdot\v')^2\over\pi^2 m^2 \sqrt{1-(\v\cdot\v')^2}}
       \left[
          - {1\over\eps}
          + \ln\left(m\over\mu\right)
          + \half\gammaE
          - {\textstyle{1\over2}}
       \right] .
\label{eq:MsqrIR2}
\end {eqnarray}


\subsubsection{The non-infrared piece}

Now turn to the remaining piece,
$\int_\q\left(|{\cal M}|^2 - |{\cal M}|_\IR^2\right)$,
which may be evaluated directly in three dimensions.
From (\ref{eq:Msqr2}) and (\ref{eq:MsqrIR1}), we have
\begin {eqnarray}
&&
   \int_\q \left(|{\cal M}|^2 {-} |{\cal M}|^2_\IR\right) =
    2\pi g^4 \!\int_{-1}^{+1} \! d\lambda \! \int_\q {1\over q}
    \Biggl\{
      \left|
          {\v\cdot\v' - \lambda^2\over q^2(1{-}\lambda^2) + \PiT(\lambda)} +
          {-1 + \lambda^2 \over q^2(1{-}\lambda^2) + \PiL(\lambda)}
      \right|^2 \!f_3(\lambda,\v\cdot\v')
\nonumber\\ && \hspace{16em} {}
        - {(\v\cdot\v')^2 \over \left| q^2 + \PiT^\IR(\lambda) \right|^2}
            \, f_3(0,\v\cdot\v')
    \Biggr\} \,.
\label{eq:Msqr3}
\end {eqnarray}
The explicit form of $f_3$ is
\begin {equation}
   f_3(\lambda,\v\cdot\v') =
   {1\over 2\pi}
   \left[1 - (\v\cdot\v')^2 - 2\lambda^2 \, (1-\v\cdot\v')\right]^{-1/2}
   .
\label {eq:f3}
\end {equation}
(This is really multiplied by a step function which vanishes
when the argument of the square root goes negative.)
The basic 3-dimensional $\q$ integral required is
\begin {equation}
   \int_\q {1\over q \, (q^2+z_1) (q^2+z_2)}
   = {\ln z_2 - \ln z_1 \over 4\pi^2 (z_2-z_1)} \, ,
\end {equation}
where the cut of the logarithm is understood to run along the negative
real axis.
It is convenient to rewrite $\Pi_{\rm L}$ and $\Pi_{\rm T}$ as
\begin {equation}
   \Pi_{\rm L,T}(\lambda) \equiv
   m^2 \, (1{-}\lambda^2)\, \rho_{\rm L,T}(\lambda) .
\end {equation}
We then obtain
\begin {eqnarray}
   \int_\q \left(|{\cal M}|^2 - |{\cal M}|^2_\IR\right)
   &=&
    {g^4\over 2\pi m^2} \int_{-1}^{+1} d\lambda \,
    \Biggl\{
    f_3(\lambda,\v\cdot\v')
    \Biggl[
          \left(\v\cdot\v' - \lambda^2 \over 1-\lambda^2\right)^2 \>
             {\arg \rhoT(\lambda) \over \Im \, \rhoT(\lambda)}
          + {\arg \rhoL(\lambda) \over \Im \, \rhoL(\lambda)}
\nonumber\\ && \qquad\qquad\qquad\qquad\qquad\quad
          - 2 \left(\v\cdot\v' - \lambda^2 \over 1-\lambda^2\right)
             \Re \, {\ln \rhoT^*(\lambda) - \ln\rhoL(\lambda) \over
                  \rhoT^*(\lambda) - \rhoL(\lambda)}
    \Biggr]
\nonumber\\ && \qquad\qquad\qquad
    - {2\over |\lambda|} \, f_3(0,\v\cdot\v') \, (\v\cdot\v')^2
    \Biggr\}\,,
\label{eq:Msqr4}
\end {eqnarray}
where $\arg z \equiv \Im \, (\ln z)$ is to be understood to lie in the range
$[-\pi,\pi]$.

Finally, we need explicit formulas for $\rhoT$ and $\rhoL$ in
three dimensions.
One can look up the formulas for $\Pi$ \cite{pi}\
or easily derive them from
Eq.~(\ref{eq:pi1}).  In the case at hand, we are interested in space-like
momenta $Q$, and the results are
\begin {mathletters}
\label {eq:rho}
\begin {eqnarray}
   \rhoL(\lambda) &=& 1 - {\lambda\over2} \ln\left( 
                                   1+\lambda\over1-\lambda \right)
                      + {i\pi\over2} \, \lambda \,,
\\
   \rhoT(\lambda) &=& {1\over2} \left[{1\over1-\lambda^2}
                                       - \rhoL(\lambda)\right] .
\end {eqnarray}
\end {mathletters}

By combining Eqs.\ (\ref{eq:dc2}), (\ref{eq:split}), (\ref{eq:MsqrIR2}),
(\ref{eq:f3}), (\ref{eq:Msqr4}), and (\ref{eq:rho}),
we now have a complete, if somewhat cumbersome and inelegant,
integral formula for $\delta C(\v,\v')$
at leading order in $g$.
Because of the remaining $\lambda$ integration, the functional dependence
of $\delta C$ on $\v\cdot\v'$ is not simple.
Fortunately, we do not need the complete form of $\delta C(\v,\v')$
to calculate the NLLO conductivity, and we will now specialize to the
calculation of the matrix element $\gamma_1$.


\subsubsection{Calculation of $\gamma_1$ and $\gamma$}

Using the general formulas (\ref{eq:dc1}) or (\ref{eq:dc2})
for $\delta C$, we have
\begin {equation}
   \gamma_1 = {\langle v^i \delta C v^i \rangle} =
   {\CA m^2 T \over 2 g^2 \mu^\eps}
          \left\langle \int_\q |{\cal M}(\v,\v',\q)|^2 \,
          (1-\v\cdot\v') \right\rangle_{\v\v'}
   .
\label {eq:gam1a}
\end {equation}
As it turns out, we can easily calculate
$\gamma$, defined by (\ref{eq:gama}), at the same time as $\gamma_1$.
So we will, even though we don't actually need the NLLO value of $\gamma$
for our calculation of the NLLO conductivity.
To this end, we define
\begin {equation}
   \gamma^{(\eta)} =
   {\CA m^2 T \over 2 g^2 \mu^\eps}
          \left\langle \int_\q |{\cal M}(\v,\v',\q)|^2 \,
          (1- \eta \, \v\cdot\v') \right\rangle_{\v\v'}
   ,
\label {eq:gamb}
\end {equation}
where $\eta = 1$ yields $\gamma_1$ and $\eta = 0$ yields $\gamma$.
We now apply this to the pieces (\ref{eq:MsqrIR2}) and (\ref{eq:Msqr4})
of $\int_\q |{\cal M}|^2$ using the three dimensional identities
\begin {eqnarray}
   \Bigl\langle f_3(\lambda,\v\cdot\v') \,
           (1-\eta\v\cdot\v')\Bigr\rangle_{\v\v'}
   &=& \Bigl\langle \delta(\lambda-\hat\q\cdot\v) \>
                  \delta(\lambda-\hat\q\cdot\v') \>
                  (1-\eta\v\cdot\v')
     \Bigr\rangle_{\hat\q\v\v'}
\nonumber\\
   &=& \fourth (1 - \eta\lambda^2) \,,
\end {eqnarray}
and similarly
\begin {eqnarray}
   \Bigl\langle (\v\cdot\v'-\lambda^2) \, f_3(\lambda,\v\cdot\v') \,
                (1-\eta\v\cdot\v')\Bigr\rangle_{\v\v'}
   &=& - \eighth \eta \, (1-\lambda^2)^2 \,,
\\
   \Bigl\langle (\v\cdot\v'-\lambda^2)^2 f_3(\lambda,\v\cdot\v') \,
                (1-\eta\v\cdot\v')\Bigr\rangle_{\v\v'}
   &=& \eighth (1-\lambda^2)^2 \, (1-\eta\lambda^2)\,,
\end {eqnarray}
plus the $d=3-\eps$ identity
\begin {eqnarray}
   \Bigl\langle (\v\cdot\v')^2 f_d(0,\v\cdot\v') \,
                (1-\eta\v\cdot\v')\Bigr\rangle_{\v\v'}
   &=& \Bigl\langle (\v\cdot\v')^2 \delta(\hat\q\cdot\v)
                \> \delta(\hat\q\cdot\v') \>
       \Bigr\rangle_{\hat\q\v\v'}
\nonumber\\
   &=& {1 \over d{-}1} \left({S_{d-2}\over S_{d-1}}\right)^2
   \, .
\end {eqnarray}
Inserting
Eq.\ (\ref{eq:MsqrIR2}) into (\ref {eq:gamb}) then gives
\begin {eqnarray}
   \gamma^{(\eta)}_\IR &=& 
   - \CA \alpha T \,
     {2 S_{d-2}^2 \over (d{-}1)\pi \, S_{d-1} \cos(\eps\pi/4)} \,
     {1\over \kappa\eps}
     \left({m \over 4\mu}\right)^{-\eps}
     \left(\pi\over\kappa\right)^{\eps/2}
\nonumber\\
   &=& \CA \alpha T \left[
          - {1\over\eps}
          + \ln\!\left(m\over\mu\right)
          + \half\gammaE
          - 2 \ln 2
          + O(\eps)
       \right] .
\end {eqnarray}   
And inserting
the non-infrared piece (\ref{eq:Msqr4})
into (\ref {eq:gamb}) now produces
\begin {equation}
   \gamma^{(\eta)} - \gamma^{(\eta)}_\IR = \CA \, \alpha T \, a^{(\eta)} ,
\end {equation}
where the numerical constant $a^{(\eta)}$ is
given by the one dimensional integral
\begin {eqnarray}
   a^{(\eta)} \equiv
   {1 \over 4}
   \int_{-1}^{+1} d\lambda \>
   \Biggl\{ &&
          (1{-}\eta\lambda^2) \,
          \left[
              {1 \over 2}\,
             {\arg \rhoT(\lambda) \over \Im \, \rhoT(\lambda)}
              + {\arg \rhoL(\lambda) \over \Im \, \rhoL(\lambda)}
          \right]
          + \eta \, {(1{-}\lambda^2)} \,
             \Re \!\left[ {\ln \rhoT^*(\lambda) - \ln\rhoL(\lambda) \over
                  \rhoT^*(\lambda) - \rhoL(\lambda)}
                  \right]
\nonumber\\ &&{}
          - {1\over |\lambda|}
    \Biggr\} \,.
\label{eq:aeta}
\end {eqnarray}
It is useful for numerical evaluation to split $\rhoT$ and $\rhoL$
into their real and imaginary parts, $\rhoT = \RT + i\IT$, {\em etc.},
and use
\begin {equation}
   {\arg \rho \over \Im \, \rho}
   = {1\over|I|} \cot^{-1}\!\left(R\over|I|\right) ,
\end {equation}
and
\begin {eqnarray}
   \Re \, {\ln \rhoT^*(\lambda) - \ln\rhoL(\lambda) \over
           \rhoT^*(\lambda) - \rhoL(\lambda)}
   &=&
       {1\over2} \,
       {\RT{-}\RL\over \left[(\RT{-}\RL)^2+(\IT{+}\IL)^2\right]} \>
       \ln\!\left(\RT^2+\IT^2 \over \RL^2+\IL^2\right)
\nonumber\\ &+&
     {|\IT{+}\IL|\over \left[(\RT{-}\RL)^2+(\IT{+}\IL)^2\right]} \,
     \left[\,
           \cot^{-1}\!\left(\RL\over|\IL|\right)
         - \cot^{-1}\!\left(\RT\over|\IT|\right)
       \right] ,
\end {eqnarray}
where
$
   \cot^{-1} x \equiv {\pi\over2} - \Tan^{-1} x
$
is defined to lie in the range $[0,\pi]$,
and we have made use of the fact that the signs of
$\IL$ and $\IL+\IT$ are the same and are both opposite to $\IT$.
One may also note that the integrand of (\ref{eq:aeta}) is
even in $\lambda$.

Putting everything together, we find
\begin {eqnarray}
   \gamma_1
   &=& \CA \, \alpha T \left[
          - {1\over\eps}
          + \ln\!\left(m\over\bar\mu\right)
          + \half \ln {\pi\over 4}
          + a_1
       \right] ,
\label{eq:gamma1}
\\
\noalign {\hbox {and}}
   \gamma
   \phantom{{}_1}
   &=& \CA \, \alpha T \left[
          - {1\over\eps}
          + \ln\!\left(m\over\bar\mu\right)
          + \half \ln {\pi\over 4}
          + a
          \phantom{{}_1}
       \right] ,
\end {eqnarray}   
where we have now written the result in terms of the \MSbar\ scale
$
   \bar\mu = \sqrt{4\pi} e^{-\gammaE/2} \mu \,.
$
Numerical evaluation of (\ref{eq:aeta})
for $\eta=1$ and $\eta=0$ gives
\begin {eqnarray}
   a_1 &=& 0.323833\cdots \, ,
\label{eq:a1}
\\
\noalign {\hbox{and}}
   a &=& 0.120782\cdots \, ,
\end {eqnarray}
for $\eta=1$ and $\eta=0$, respectively.
In fact, our numerical evaluation of the constant $a$ shows that it
precisely equals $-\half \ln {\pi\over 4}$ to 12 significant digits.
Surely this is an exact identity,%
\footnote
    {
    We have not bothered to try proving this analytically,
    although we are confidant one could do so.
    Blaizot and Iancu \cite {BlaizotGamma} have shown the corresponding result
    when using a sharp momentum IR cutoff,
    instead of dimensional regularization.
    (See appendix~B of \cite {block-nordsieck}.)
    }
so that the dimensionally regulated hard gauge boson damping rate,
to next-to-leading-log order, is simply
\begin {equation}
   \gamma = \CA \, \alpha T \left[
          - {1\over\eps}
          + \ln\!\left(m\over\bar\mu\right)
	  \right] .
\end {equation}


\subsection {Matching to Theory 2}

In theory 2, the total effective collision term is, in principal, composed
of two parts.  First, there's the bare collision term that appears in
(\ref{eq:2a}), which we will call $\deltaC_\bare$ here to be explicit,
and which conceptually represents collisions due to the exchange of
virtual gauge bosons that were integrated out in going from Theory 1
to Theory 2.
Secondly, there is a dynamical contribution to $\deltaC$, which we will
call $\deltaC_\dyn$, which arises from the exchange of those gauge bosons
that have {\it not}\/ yet been integrated out.
However, as we explain below, the same nice property of dimensional
regularization which
simplified the Theory 3 matching calculation in section \ref{sec:final23match}
works
here as well: for the purposes of matching, $\deltaC_\dyn$ must vanish
in dimensional regularization by dimensional analysis.
Hence, we have simply
$\deltaC_\tot = \deltaC_\bare$, and so the $\gamma_1$ we needed in order
to match Theory 2 to Theory 3 can simply be taken directly from the
Theory 1 result (\ref{eq:gamma1}) for $\langle1|\deltaC_\tot|1\rangle$.

The dimensional argument can be made by rescaling to variables
$\bar t = m^{-2} t$, $\bar\A = T^{-1/2}\A$,
$\bar A_0 =  m^2 T^{-1/2} A_0$, $\bar W = m^2 T^{-1/2} \, W$,
and $\bar\xi = m^2 T^{-1/2} \, \xi$, so that
the equations (\ref{eq:2}) for Theory 2 become
\begin {mathletters}
\label {eq:2rescale}
\begin {eqnarray}
&
   \v\cdot\bar\D \, \bar W - \v\cdot\bar\E
   =
   - \deltaC_\bare \,\bar W + \bar\xi ,
&\\&
   \langle \bar W \rangle = 0 ,
&\\&
   \bar\D \times \bar\B = \langle \v \bar W \rangle ,
&\\&
   \dlangle \bar \xi(\x,\v,t) \, \bar \xi(\x',\v',t') \drangle =
   2 \, \deltaC_\bare(\v,\v') \, \delta(\x{-}\x') \, \delta(t{-}t') ,
&
\end {eqnarray}
\end {mathletters}%
where $\bar D_\nu = \bar\partial_\nu + g \mu^{\eps/2} T^{1/2} \bar A_\nu$.
There remains a dimensional quantity, $\deltaC$, other than the
effective coupling $g \mu^{\eps/2} T^{1/2}$.
But in matching theories 1 and 2, it is important that
$\deltaC_\bare$ is to be formally treated as a perturbation.
This is feasible because
the matching is performed at the spatial momentum scale $m = O(gT)$,
whereas $\deltaC_\bare$
is of the much more infrared scale $\gamma = O(g^2 T \ln)$.
As discussed earlier, matching can always be thought of as taking place in
a large box, serving as an infrared regulator.  The box should be chosen
to be large compared to the distance scale of the matching ($1/m$), but
may be small compared to more infrared scales where the physics
becomes more complicated ({\em e.g.}, $1/\gamma$).
In the presence of such an
IR cut-off, one may then treat quantities that are soft relative to the
matching scale (such as $\deltaC_\bare$, in the case at hand) as perturbations.
Moreover, formally treating them as perturbations works for the
purposes of matching calculations even if the infrared regulator is then
taken to arbitrarily large distance scales \cite{Braaten&Nieto,dimreg}.

In summary, then, perturbation theory in the effective coupling
$g \mu^{\eps/2} T^{1/2}$ and in $\deltaC$ can only give terms with
the dimensions of
$\left(g \mu^{\eps/2} T^{1/2}\right)^m \left(\deltaC\right)^n$
for integer $m$ and $n$.
Because of the factor of $\mu^\eps$,
none of these can consistently match the mass dimension 1 of
$\deltaC_\tot$ unless $m=0$.  That means the only
contribution to $\deltaC_{\tot}$ is the tree-level
$\deltaC_\bare$.

Though the dynamical contributions to $\deltaC$ formally vanish in Theory 2 in
dimensional regularization, they do play an important conceptual role.
We are now interpreting Eq.~(\ref{eq:gamma1}) as the result for the bare
$\gamma_1$ in Theory 2.  Theory 2 requires ultraviolet regularization:
the $1/\eps$ in (\ref{eq:gamma1}) is the counter-term for a UV
divergence in Theory 2, and $\mu$ is the associated renormalization scale.
In our matching calculation, however, the $1/\eps$ and the
$\ln(m/\mu)$ actually arose in Eq.~(\ref{eq:MsqrIR2}) from a formal
{\it infrared}\/ divergence of
the calculation of the total $\gamma_1$ in Theory 1.
The discrepancy of interpretation is resolved by realizing that the
Theory 2 result that $\gamma_{1,\tot} = \gamma_{1,\bare}$ should really be
thought of as
\begin {equation}
   \left[\gamma_{1,\tot}\right]_{\rm(Theory 2)}
   = \gamma_{1,\bare}
     + \CA\alpha T \left\{
          \left[ -{1\over\eps} + \ln\left(m \over\mu\right) \right]_{\rm IR}
        + \left[ +{1\over\eps} - \ln\left(m \over\mu\right) \right]_{\rm UV}
     \right\} ,
\label {eq:dimreg}
\end {equation}
and so equating the total $\gamma_1$ in the two theories converts the
the IR divergence into a UV divergence in (\ref{eq:gamma1}).
Eq.\ (\ref{eq:dimreg}) is an example of the generic behavior
in dimensional regularization of logarithmic divergences
when there is no scale to cut them off in
either the IR or UV, and is typified by the simple example
\begin {equation}
   \int {d^{3-\eps}p \over p^3}
   \propto \mu^\eps \int_0^\infty {dp \over p^{1+\eps}}
   = \int_0^\Lambda {dp \over p^{1+\eps}}
   + \int_\Lambda^\infty {dp \over p^{1+\eps}}
   = \left[ - {1\over\eps} + \ln {\Lambda\over\mu} \right]
   + \left[ + {1\over\eps} - \ln {\Lambda\over\mu} \right] ,
\end {equation}
where in the last equality dimensional regularization was used both for the
IR contribution of the first term and the UV contribution of the second.



\section {Final Results}
\label {sec:final}

We now put together our NLLO result (\ref{eq:fsigma})
for the color conductivity $\sigma$ in terms of $\gamma_1$
and our result (\ref{eq:gamma1}) for $\gamma_1$.
The structure is clearest if we write an expansion for $\sigma^{-1}$
(the ``color resistivity'')
rather than $\sigma$ directly.
One finds
\begin {mathletters}
\begin {eqnarray}
   \sigma^{-1}
   &=& {3\CA\alpha T\over m^2} \left[\,
       \ln\left(m\over\gamma(\mu)\right)
          + C
          + O(\ln^{-1})
       \right] ,
\label{eq:final}
\end {eqnarray}
with
\begin {eqnarray}
    C &\equiv&
          \half \ln {\pi\over4}
          + a_1
          + I(1)
    =
          3.0410\cdots \,,
\end {eqnarray}
\label {eq:final2}%
\end {mathletters}%
and where the numerical constants $I(1)$ and $a_1$ are defined by
(\ref{eq:I}) and (\ref{eq:aeta}) [with $\eta{=}1$] and given numerically by
(\ref{eq:Inumber}) and (\ref{eq:a1}).
Note that the $1/\eps$ divergences have canceled, as they must.
Inside the logarithm of (\ref {eq:final}),
$\gamma(\mu)$ is to be understood as simply the leading-log formula
\begin {equation}
   \gamma(\mu) \approx \CA \alpha T \ln\left(m\over\mu\right) ,
\end {equation}
and $\mu$ should be chosen so that it is of order $\gamma$.
One may easily verify that the $\mu$ dependence in the NLLO result
(\ref{eq:final}) only affects that
answer at order $[\ln(m/\gamma)]^{-1} \sim [\ln(1/g)]^{-1}$, which is beyond
the order of this calculation.

Eq.\ (\ref{eq:final2}) is our final result.
Although the result must be gauge independent,
our derivation has been restricted to a particular choice of gauge,
namely Coulomb gauge.
It would be comforting to have the calculation repeated in another gauge,
perhaps the generalized flow gauges discussed in appendix \ref{app:flow},
but this we have not attempted to do.
It would also be interesting if there were any way to express the
fundamental functions $\Sigma_m(\rho)$ of this problem in terms of
standard mathematical functions, but we have been unable to do so.

Our result for the NLLO color conductivity may be compared against
numerical simulation of the electroweak baryon number violation rate,
as such simulations can in fact be used to measure the relative size
of the NLLO correction to $\sigma$.
We discuss this comparison in ref.\ \cite{overview}.

Finally, we should mention the differences between our calculation of
the NLLO conductivity and
that outlined in earlier work by Blaizot and Iancu \cite{BlaizotGamma}.
In our language, Blaizot and Iancu's discussion amounts to specifying
how to calculate $\gamma_1$ (which they call $\gamma + \delta$).
That is, it is equivalent to our discussion of matching Theory 1 and
Theory 2, though they do not actually push through the calculation to
get a final, regulated, numerical result like (\ref{eq:gamma1}).
They then quote a result for the conductivity of simply $m^2 / (3\gamma_1)$.
However, this misses all the contributions that in our calculation came
from integrating out physics at $k \sim \gamma$ when matching Theory 2
to Theory 3.  Moreover, the result $m^2/(3\gamma_1)$ is not a well-defined
quantity at NLLO because of the infrared divergence in the calculation of
$\gamma_1$.  If one simply cuts off that divergence at $k \sim \gamma$,
as suggested by Blaizot and Iancu, $\gamma_1$ is still sensitive at NLLO
to one's convention in choosing whether to cut it off at $k = \gamma$,
$k = 2 \gamma$, or something similar.  The calculation of the effects
treated in our Theory 2 to Theory 3 matching is crucial to obtain an
answer that is independent at NLLO to the precise choice of cut-off scale
(renormalization scale) $\mu$.


\section* {ACKNOWLEDGMENTS}

We are indebted to Guy Moore for useful discussions concerning the
evaluation of Wilson loops in B\"odeker's effective theory,
and we thank Dietrich B\"odeker and Edmond Iancu for a
variety of helpful discussions.
This work was supported, in part, by the U.S. Department
of Energy under Grant Nos.~DE-FG03-96ER40956
and DE-FG02-97ER41027.

\newpage

\appendix

\section{Wilson loops in flow gauges}
\label{app:flow}

A useful class of gauges for stochastic gauge theory are the flow
gauges defined by the condition
$\sigma A_0 = -\lambda \grad\cdot\A$.
This class of gauges interpolate
between $A_0=0$ gauge ($\lambda \to 0$) and Coulomb gauge
($\lambda\to\infty$)  \cite{flow gauges}.%
\footnote{
  The flow gauges here correspond to the $\xi\to\infty$ limit of the
  generalized flow gauges considered in Ref.~\cite{flow gauges}.
}
In this appendix, we will illustrate the use of these gauges by
explicitly checking the gauge-invariance of the derivation in
section \ref{sec:Wilson1} of the
the first-order result for large-time rectangular Wilson loops.

The gauge-fixed path integral, analogous to (\ref{eq:Zcoulomb}), is
\begin {mathletters}
\begin {equation}
   Z_{\rm flow} = \int [{\cal D}A_0] [{\cal D}\A]
          [{\cal D}\bar\eta] [{\cal D} \eta]
          \> \delta(\sigma A_0 {-} \lambda\grad\cdot\A) \>
          \exp\left(-\int dt \> d^3 x \> L_{\rm flow}\right) ,
\end {equation}
with
\begin {equation}
   L_{\rm flow} = {1\over 4\sigma T} \left\{
          \left| -\sigma\E + \D\times\B \right|^2
          + \bar\eta (\sigma D_0 + \lambda\grad\cdot\D) \eta
   \right\} .
\end {equation}
\end {mathletters}%
Rewriting $A_0$ in terms of $\grad\cdot\A$ and working in momentum space,
the perturbative expansion of the action gives
\begin {equation}
   S_{\rm flow} = \int dt \int_\k \> \left[ {1\over 4\sigma T} \,
          \left|\left(\sigma \partial_t - \nabla^2\right) \A
                + (1{-}\lambda) \grad \grad\cdot\A
          \right|^2
          + O(\A^3)
          + \hbox{(ghosts)}
   \right] .
\end {equation}
The propagator is
\begin {equation}
   A^a_i \wiggle A^b_j
   = 2\sigma T \delta^{ab} \left[
         {\delta_{ij}-\hat k_i \hat k_j \over
                    \left|i\sigma\omega + k^2\right|^2}
         + {\hat k_i \hat k_j \over
                    \left|i\sigma\omega + \lambda k^2\right|^2}
     \right] .
\label{eq:flow1}
\end {equation}
One can see that the gauge $\lambda=1$ is the stochastic gauge theory analog
of Feynman gauge.  This Feynman-like gauge was first proposed in this context
by Zwanziger \cite{Zwanziger}.

In any case, let's now turn to the Wilson loop diagrams of
Fig.\ \ref{fig:diags}c.
We need the propagator for $A_0 = (\lambda/\sigma) \grad\cdot\A$ which, from
(\ref{eq:flow1}), is
\begin {equation}
   A_0^a \wiggle A_0^b
   = {2T \over \sigma} \, \delta^{ab} \,
         {\lambda^2 k^2 \over \left|i\sigma\omega + \lambda k^2\right|^2}
     \, .
\end {equation}
Then, following (\ref{eq:Wilson1}), 
\begin {eqnarray}
   d_\R^{-1} \, \delta \Wil
   &=& - {g^2 \tr(T^a T^b) \over \tr(1)} \int_0^{\ttot} dt\>dt'\>
        \Bigdlangle A_0^a(t,0) A_0^b(t',\vecR) \Bigdrangle
\nonumber\\
   &=& - g^2 \CA \ttot \int_\k
         \left[ {2 T \over \sigma} \,
            {\lambda^2 k^2 \over \left|i\sigma\omega + \lambda k^2\right|^2}
         \right]_{\omega{=}0}
         e^{i\k\cdot\vecR}
   = - {\alpha \CA T \over \sigma R} \, \ttot \,,
\end {eqnarray}
in the large-time limit.
As required for any physical quantity,
all dependence on the gauge-fixing parameter $\lambda$
has disappeared.


\section{Large \boldmath$R$ behavior of Wilson loops}
\label{app:Wilson}

Consider the first-order correction to the real-time Wilson loop expectation
(\ref{eq:dWilson1}), repeated here for convenience,
\begin {equation}
   d_\R^{-1} \, \delta \Wil
   = - 2 g^2 T\, \CA \ttot
   \int_\k {e^{i\k\cdot\vecR} \over k^2 \, \sigmaL(k) } \, ,
\label {eq:dW2}
\end {equation}
in the limit of large $R$.
Specifically $R$ must be large compared to the
color-changing mean free path $\gamma^{-1} = O\left[(g^2 T\ln)^{-1}\right]$.
Physically, one should simultaneously keep $R$
sufficiently small that physics on the scale $R$ is still perturbative,
that is, $R \ll (g^2 T)^{-1}$.
Formally, when performing an IR regulated matching calculation,
one may simply take $R \to \infty$.
But here, we will examine what happens if one is not quite so cavalier with $R$.
For the sake of definiteness, we will consider $R \sim \delta (g^2 T)^{-1}$,
where $1/\ln g^{-1} \ll \delta \ll 1$ and the coupling $g$ is arbitrarily weak.

Rewrite the Fourier transform in (\ref{eq:dW2}) as
\begin {equation}
   {\cal F}(R) \equiv \int_\k {s(k) \over k^2} \, e^{i\k\cdot\R} ,
\end {equation}
where $s(k) \equiv [\sigmaL(k)]^{-1}$, and perform the angular integral:
\begin {equation}
   {\cal F}(R) = {1\over 2\pi^2 R}
      \int_0^\infty {dk \over k} \, s(k) \, \sin(kR) \,.
\label {eq:F}
\end {equation}
The analytic continuation of $s(k)$ is an even function of $k$, as can
be verified explicitly from the formulas (\ref{eq:sLG0}) and (\ref{eq:cf}).
So (\ref {eq:F}) can be rewritten as
\begin {eqnarray}
   {\cal F}(R)
   &=& \lim_{\varepsilon\to 0} \, {1\over 4\pi^2 i R}
      \int_\varepsilon^\infty dk\> {s(k) \over k} \,
      \Bigl[e^{ikR} - e^{-ikR} \Bigr]
   = {1\over 4\pi^2 i R}
      \int_{-\infty}^\infty dk \>
      e^{ikR} \> \mbox {P.P.} \biggl({s(k)\over k}\biggr) \,.
\label {eq:foop}
\end {eqnarray}
Here, P.P.~denotes principal part,
\begin {equation}
   \mbox{P.P.} \left({1\over k}\right)
   = {1\over2} \left({1\over k-i\varepsilon}
   + {1\over k+i\varepsilon}\right) .
\end {equation}
Closing the contour of (\ref{eq:foop}) in the upper half plane then
picks up the $k = i\varepsilon$ pole, as well as any contributions from
singularities in the function $s(k)$.  The $k = i\varepsilon$ pole
gives the expected result (\ref{eq:Wilson1b}).
Singularities in $s(k) = [\sigmaL(k)]^{-1}$ 
at some $k$ in the upper half plane
will give additional contributions suppressed by $\exp[-R \, \Im(k)]$.
$\sigmaL(k)$ is analytic and non-vanishing in a neighborhood of the real axis.
The only scale associated with the $k$ dependence
of $\sigmaL(k)$ is the scale $\gamma$ appearing in $\deltaC$.
So the only scale that can determine the imaginary parts of the singularities
of $s$ is $\gamma$.
Therefore,
the contribution to (\ref{eq:foop}) from singularities of $s$ must
be suppressed by $\exp[-O(\gamma R)]$.
For $R \sim \delta (g^2 T)^{-1}$ this is $\exp [-\delta \, O(\ln g^{-1})]$
which, for weak coupling, behaves as some (positive) power $g$ and 
vanishes faster than any power of $1/\ln g^{-1}$.


\section{Relations between various correlators}
\label{app:props}

Consider, for simplicity,
an effective theory for a
single real scalar field,
governed by the classical Langevin equation
$\sigma \dot\phi = (\grad^2{-}m^2)\, \phi - {d V_{\rm int}/d\phi} + \zeta$
and the noise covariance $\dlangle \zeta \zeta \drangle = 2\sigma T$.
Also for simplicity, take $\sigma$ to be constant.
The corresponding action in a path integral formulation is then%
\footnote
    {%
    Formally,
    the action should also contain the Jacobian term
    $
	-\half \, \tr \!\left(-\grad^2 + m^2 + V_{\rm int}''(\phi) \right)
    $.
    But this is proportional to $\delta^{(d)}(0)$ for local potentials
    $V_{\rm int}(\phi)$ and so vanishes in dimensional regularization.
    See, for example, sec.\ 17.2 of ref.\ \cite{ZinnJustin}.
    }
\begin {equation}
   S = {1\over 4\sigma T} \int dt\> d^3x
           \left[ \sigma \dot\phi - (\grad^2{-}m^2) \,\phi
                - {d V_{\rm int}\over d\phi}
           \right]^2 .
\label {eq:L}
\end {equation}
If $\delta V/\delta\phi$ is treated as a perturbation, then
the unperturbed propagator arising from this action is, in Fourier space,
\begin {equation}
   \dlangle \phi(\omega,\k) \, \phi(\omega',\k')^* \drangle
   = -i \tilde G(\omega,\k) \>
     (2\pi)^4 \delta(\k{-}\k') \> \delta(\omega{-}\omega') \,.
\end {equation}
with
\begin {equation}
   -i \tilde G(\omega,\k) = {2 \sigma T \over (\sigma \omega)^2 + (k^2+m^2)^2} \,
\label{eq:phiprop}
\end {equation}
In the main text, the term ``propagator'' referred what we've called
$-i \tilde G$ above, which in coordinate space is
$\dlangle \phi(t,\x) \, \phi(0,0) \drangle$.
In the underlying quantum field theory, however, it is customary to
define the propagator $G$ as $i$ times the expectation of a product
of fields, as indicated above.
Note from (\ref{eq:phiprop}) that $G$ is purely imaginary.

Because
operators commute in a classical theory,
there is no distinction between time-ordered and
time-unordered propagators.
So the above propagator can be regarded as
the classical limit of the time-ordered propagator%
\footnote{
   In our specific example, $\phi$ is real field,
   and so $\phi^\dagger(t,\x)$ is no
   different from $\phi(t,\x)$.
   In writing general quantum expressions for propagators and spectral
   densities and so forth, we nonetheless find it useful for the sake
   of reference to include daggers where they would appear for a
   discussion of complex fields.
}
$
    G(t,\x) =
    i\langle {\cal T} \!\{\phi (t,\x) \, \phi^\dagger (0,0) \} \rangle 
$,
or either of the
Wightman functions
$
    G_>(t,\x) =
    i\langle \phi (t,\x) \, \phi^\dagger (0,0) \rangle
$ or
$
    G_<(t,\x) =
    i\langle \phi^\dagger (0,0) \, \phi (t,\x) \rangle
$,
of the underlying quantum theory.
That is,
\begin {equation}
   G = G_> = G_< \qquad \hbox{(classically)} .
\end {equation}


A classical effective theory, such as the Langevin theory (\ref {eq:L}),
is (at best) valid for low frequency, long wavelength dynamics.
It is only appropriate for studying observables which are 
smeared over time and spatial scales large compared to $\beta\hbar$,
and thus insensitive to short time (or short distance) quantum fluctuations.

In the underlying quantum theory,
the spectral density $ \rho(\omega,\k)$
(defined as the Fourier transform of
$
    \langle [\phi (t,\x), \phi^\dagger(0,0)] \rangle
$)
is related to the time-ordered
propagator $G$ by
\begin {equation}
   \rho(\omega,\k) = 2 \tanh\left(\half\beta\hbar\omega\right) \,
                      \Im \, \tilde G(\omega,\k) \,.
\label{eq:moosepoop2}
\end {equation}
Although it is not essential to the discussion, we will keep track of factors
of $\hbar$ in this appendix (and this appendix only).
Taking the classical limit $\hbar\omega \ll T$, in which
the real-time propagator $G$ is purely imaginary, 
(\ref{eq:moosepoop2}) gives
\begin {equation}
   i\rho(\omega,\k) = \beta\hbar\omega \> \tilde G(\omega,\k) \,.
\end {equation}
The Fourier transform of the retarded propagator
$
    G_{\rm R}(t,\x) =
    i\theta(t)\> \langle [\phi(t,\x),\phi^\dagger(0,0)] \rangle
$
is related to $\rho$ by the spectral representation
\begin {equation}
   \tilde G_{\rm R}(\omega,\k) = \int_{-\infty}^{+\infty} {d\omega'\over2\pi} \>
        {\rho(\omega',\k) \over \omega'-\omega-i\varepsilon} 
    \,,
\label {eq:spec}
\end {equation}
which, in the classical limit, becomes just
\begin {equation}
   G_{\rm R}(t,\k) = i\theta(t) \, \beta\hbar \, {\partial\over\partial t} \,
                      G(t,\k) \,.
\label{eq:GR}
\end {equation}
Similarly, the advanced propagator
$
    G_A(t,\x) = -i\theta(-t)\> \langle [\phi(t,\x),\phi^\dagger(0,0)] \rangle
$
satisfies the same spectral representation (\ref {eq:spec})
except for changing $-i\epsilon$ to $+i\epsilon$,
and in the classical limit is
\begin {equation}
   G_{\rm A}(t,\k) = - i\theta(-t) \, \beta\hbar \, {\partial\over\partial t} \,
                      G(t,\k) \,.
\label{eq:GA}
\end {equation}
Note that these propagators satisfy the standard relation
$\tilde G_{\rm R} - \tilde G_{\rm A} = i\rho$.

In studying static equilibrium physics, one typically works in imaginary
time $\tau$ rather than real time $t$, where Euclidean correlation functions
such as
$
    G_{\rm E}(\tau,\x) \equiv
    \langle {\cal T} \!\{\phi (-i\tau,\x) \phi^\dagger (0,0) \} \rangle 
$
are periodic with period $\beta\hbar$.
In the classical limit, which is the long-distance limit
$\Delta x \gg \beta\hbar$, there is no substantive
difference between equal-time correlations and zero-frequency correlations in
imaginary time, due to the negligible extent $\beta\hbar$ of imaginary time.
In this limit, the only discrepancy is a change in normalization of
$\beta\hbar$ from the Fourier transform,
\begin {equation}
   G_{\rm E}(\tau{=}0, \k)
   = (\beta\hbar)^{-1} \, \tilde G_{\rm E}(\nu{=}0,\k) \,.
\end {equation}
The limit of zero time separation is the same in imaginary or real time,
so one must have $G(t{=}0,\k) = iG_{\rm E}(\tau{=}0,\k)$, since
the real-time time-ordered propagator is related to the Euclidean propagator
by appropriate analytic continuation in $t$.
But the analytic continuation in {\em frequency} of $\tilde G_E$
from the Matsubara points $\nu_n = 2\pi i n / (\beta\hbar)$
back to real frequencies
yields $\tilde G_{\rm R}$ or $\tilde G_{\rm A}$,
depending on whether one continues to the real frequency axis
from above or below, respectively.
Hence, the zero frequency retarded, advanced, and Euclidean
propagators coincide,
$
      \tilde G_{\rm E}(\nu{=}0,\k)
    = \tilde G_{\rm R}(\omega{=}0,\k)
    = \tilde G_{\rm A}(\omega{=}0,\k)
$.
Putting everything together gives the following string of
equalities in the classical limit:
\begin {eqnarray}
    G_{\rm E}(\tau{=}0,\k)
    &=& (\beta\hbar)^{-1} \, \tilde G_{\rm E}(\nu{=}0,\k)
    = (\beta\hbar)^{-1} \, \tilde G_{\rm R}(\omega{=}0,\k)
    = (\beta\hbar)^{-1} \, \tilde G_{\rm A}(\omega{=}0,\k)
\nonumber\\ &=&
    -iG(t{=}0,\k)
    = -i\int_{-\infty}^{+\infty} {d\omega\over 2\pi} \> \tilde G(\omega,\k)
    \,.
\label {eq:chain}
\end {eqnarray}
Note that this also implies
\begin {equation}
   G(t{=}0,\k)
   = i(\beta\hbar)^{-1} \int_{-\infty}^{+\infty} dt \> G_{\rm R}(t,\k)
   = i(\beta\hbar)^{-1} \int_{-\infty}^{+\infty} dt \> G_{\rm A}(t,\k) ,
\end {equation}
which can be seen directly from (\ref{eq:GR}) and (\ref{eq:GA}).

It is important to understand that the small frequency limit
of the real-time correlator $\tilde G(\omega,\k)$ is {\em not}
directly related to the zero frequency limit of the Euclidean correlator
$\tilde G_{\rm E}(\nu,\k)$.
As an example, for our real-time propagator (\ref{eq:phiprop}),
\begin {eqnarray}
   \tilde G(\omega{=}0,\k) &=& {2 i\sigma T \over (k^2+m^2)^2} \,,
\\
\noalign {\hbox {while}}
   G(t{=}0,\k) &=& \int_{-\infty}^{+\infty} d\omega \> G(\omega,\k)
            = {i T \over k^2{+}m^2} \,.
\end {eqnarray}
In this simple scalar model,
the chain of relations (\ref {eq:chain}) imply that
$
   G_{\rm E}(\tau{=}0, \k)
$
and
$
   (\beta\hbar)^{-1} \, \tilde G_{\rm E}(\nu{=}0,\k)
$
likewise equal $ T / (k^2+m^2) $.
Some additional specific results for this model are
\begin {eqnarray}
   \rho(\omega,\k)
   &=& {2\sigma\hbar\omega \over (\sigma\omega)^2 + (k^2+m^2)^2} \,,
\label {eq:specden}
\\
   \tilde G_{\rm R}(\omega,\k)
   &=& {\hbar \over -i\sigma\omega + k^2 + m^2} \,,
\\
   \tilde G_{\rm A}(\omega,\k)
   &=& {\hbar \over i\sigma\omega + k^2 + m^2} \,.
\end {eqnarray}
Finally, if one inserts the spectral density (\ref {eq:specden}) directly
into the spectral representation, in the underlying quantum theory,
of the Euclidean propagator
\begin {equation}
   \tilde G_{\rm E}(\nu_n,\k) = \int_{-\infty}^{+\infty} {d\omega'\over2\pi} \>
        {\rho(\omega',\k) \over \omega' + i\nu_n} 
    \,,
\end {equation}
then one obtains
\begin {equation}
   \tilde G_{\rm E}(\nu_n,\k)
   = {\hbar \over \sigma|\nu_n| + k^2 + m^2} \,.
\label {eq:crazy}
\end {equation}
However, this result is not sensible (for non-zero Matsubara frequencies).
As noted earlier, the classical Langevin theory which led to
(\ref {eq:specden}) is only valid for frequencies small compared
to $(\beta\hbar)^{-1}$, whereas the result (\ref {eq:crazy})
depends sensitively on the precise form of the high frequency
behavior of the spectral density (\ref {eq:specden}).


\section{Small $\p$ expansion of \boldmath$\Gop$}
\label{app:G0x}

In this appendix, we discuss how to expand
$\Gop = [\v\cdot i\p + \deltaC]^{-1}$ for small momentum $\p$.
We cannot simply treat $i\v\cdot\p$ as a perturbation
because $\deltaC$ has a zero mode (\ref{eq:dC0}) and so is not invertible.
The correct way to expand is to separate $i\v\cdot\p$
into zero-mode and non-zero mode pieces,
\begin {equation}
    \v\cdot i\p =
    \v\cdot i\p \, \Po + \Po \, \v\cdot i\p
    + (1{-}\Po) \, \v\cdot i\p \, (1{-}\Po) \,,
\label {eq:convective}
\end {equation}
and treat only the last term as a perturbation \cite{Blog2}.
The first two terms are rank-one operators which will lift the zero mode of
$\deltaC$.
This leads to the expansion
\begin {mathletters}
\label{eq:G0x1}
\begin {eqnarray}
   \Go &=& \left[\Goo^{-1} + (1{-}\Po) \, \v\cdot i\p \, (1{-}\Po)\right]^{-1}
\nonumber\\
       &=& \Goo
         - \Goo \, (1{-}\Po) \, \v\cdot i\p \, (1{-}\Po) \, \Goo
\nonumber\\ && \hspace{1.9em} {}
         + \Goo \, (1{-}\Po) \, \v\cdot i\p \, (1{-}\Po) \, \Goo
                     \, (1{-}\Po) \, \v\cdot i\p \, (1{-}\Po) \, \Goo
         - \cdots ,
\label{eq:G0x1a}
\end {eqnarray}
where
\begin {eqnarray}
   \Goo &\equiv&
   \left[ \v\cdot i\p \, \Po + \Po \, \v\cdot i\p + \deltaC \right]^{-1}
\nonumber\\
   &=& (1{-}\Po) \, \deltaC^{-1} \, (1{-}\Po)
             +  {d \over \gamma_1 p^2} \, (\gamma_1 - \v\cdot i\p) \, \Po \,
                                          (\gamma_1 - \v\cdot i\p) .
\end {eqnarray}
\end {mathletters}%
To verify the last equality, note that
$\deltaC \, \v\cdot i\p \Po = \gamma_1 \v\cdot i\p \Po$ by (\ref{eq:dC1}).
Observe that $\Goo$ is $O(p^{-2})$.
That might appear problematical for the expansion (\ref{eq:G0x1a}), which
brings along a factor of $\Goo$ with every factor of $\v\cdot i\p$,
except that the inner factors of $\Goo$ always appear in the
combination
\begin {equation}
   (1{-}\Po) \, \Goo \, (1{-}\Po) =
   (1{-}\Po) \left[ \deltaC^{-1}
             -  {d \over \gamma_1} \, \v\cdot i\hat\p \, \Po \, \v\cdot i\hat\p
             \right] (1{-}\Po) ,
\end {equation}
which is only $O(p^0)$.

The expansion (\ref{eq:G0x1}) yields
\begin {equation}
   \Go = {d\over\gamma_1 p^2} \, (\gamma_1-\v\cdot i\p) \,
                       \Po \, (\gamma_1 - \v\cdot i\p) + O(p^0) ,
\label {eq:G0x2}
\end {equation}
and so
\begin {equation}
   \Go \rangle = {d\over p^2} \, (\gamma_1-\v\cdot i\p) \Bigr\rangle + O(p^0)
   \,,
\end {equation}
and
\begin {equation}
   \langle \Go \rangle = {d\gamma_1\over p^2} + O(p^0) \,.
\end {equation}
Putting the last two equations together gives the expansion
(\ref{eq:GG0x}) for $\GGo\rangle$ cited in the main text.

The expansion (\ref{eq:G0x1}) also gives
\begin {equation}
   \sigma_\p = {m^2\over d{-}1} \, \langle v_i \Go v_i \rangle
             = \sigma_0  + O(p^2) \,,
\end {equation}
which is a result used in section (\ref{sec:IR}).


\section{Large $\p$ expansion of \lowercase{\boldmath$\Sigma_m(\rho)$}}
\label{app:largep}

Using the large $\p$ expansion (\ref{eq:GoUV}) of $\Gop$, one finds
\begin {eqnarray}
   \Sigma_m(\rho) &\sim&
   \langle mm | {\rm P.P.} \, {1\over i v_z \rho} + \pi\delta(v_z \rho)
           | mm \rangle
   = {\pi\over\rho} \, \langle mm | \, \delta(v_z) \, | mm \rangle
\nonumber\\
   &=& {(2m{+}1)\over(2m)!} \left[ P_m^m(0) \right]^2  {\pi\over2\rho}
   = {(2m{+}1)!! \, (2m{-}1)!!\over (2m)!} \, {\pi\over 2\rho} \,,
\end {eqnarray}
where $(-1)!! \equiv 1$.
[The P.P. terms vanishes because the $|mm\rangle$ states are invariant
under $v_z \to -v_z$.]

We can make a little progress analyzing the corrections to this leading term
by attempting to treat $\deltac$ as a perturbation in
$(iv_x\rho + \varepsilon)^{-1}$,
\begin {equation}
   \Sigma_{m}(\rho) =
   \langle mm | (i v_z \rho + \varepsilon)^{-1} | mm \rangle
   - \langle mm | (i v_z \rho + \varepsilon)^{-1} \, \deltac \,
                  (i v_z \rho + \varepsilon)^{-1} | mm \rangle
   + \cdots
\end {equation}
Call the second term of the expansion $\delta\Sigma_m$.  Taking the
formula (\ref{eq:deltac}) for $\delta c$, we can rewrite this term as
\begin {eqnarray}
&&
   \delta\Sigma_{m}(\rho) =
   \langle mm | (i v_z \rho +\varepsilon)^{-2} | mm \rangle
\nonumber\\ && \quad
   - 4\pi \left\langle Y_{mm}^*(\v) \,
        \left[{\rm P.P.} \, {1\over i v_z \rho} + \pi\delta(v_z \rho)\right]
        \, \deltac_2(\v,\v') \,
        \left[{\rm P.P.} \, {1\over i v'_z \rho} + \pi\delta(v'_z \rho)\right]
        Y_{mm}(\v')
     \right\rangle_{\v\v'} ,
\label {eq:dSig}
\end {eqnarray}
where
\begin {equation}
   \delta c_2(\v,\v') \equiv
   - {4\over\pi} {(\v\cdot\v')^2\over \sqrt{1-(\v\cdot\v')^2}} \, .
\end {equation}
By independently considering both $\v\to-\v$ and $\v'\to-\v'$ in the
second term of (\ref{eq:dSig}), one sees that
only the
$\delta$-functions will contribute for even $m$ and only the
principal-part for odd $m$.

As we shall see, the $\v\v'$ average in the second term of
(\ref{eq:dSig}) is logarithmically
divergent as simultaneously $v_z \to 0$ and $\v' \to \pm \v$:
a divergence which the
$\varepsilon$ prescription does not cure.
In contrast, the first term of (\ref{eq:dSig}) does not give rise to
a logarithm.  For example, for $m=0$, one gets
\begin {equation}
   \langle 00 | (i v_z \rho + \varepsilon)^{-2} | 00\rangle
   = {1\over2} \int_{-1}^{+1} {dv_z \over (i v_z\rho + \epsilon)^2}
   = {1\over p^2} \,.
\end {equation}
We will now focus on the logarithmic divergence and so ignore the
first term of (\ref{eq:dSig}).
The divergence is cut off only
by $\deltaC$ itself.  That is, it is cut off when $|i v_z \rho|$ becomes
of order $\deltac = O(1)$, so that $|v_z| \sim \rho^{-1}$.  The
logarithmic divergence of (\ref{eq:dSig}) is an artifact of our treatment of
$\deltac$ as a perturbation.  However, by proceeding with (\ref{eq:dSig})
and treating $v_z$ as effectively cut off at $O(\rho^{-1})$, and so
$\delta(v_z\rho)$ as being smeared to have a width of
$\Delta(v_z\rho) \sim 1$, we may extract the coefficient of the logarithm.

The log divergence can be seen by setting $v_z = v_z' = 0$ and considering
just the integral over azimuthal angles in (\ref{eq:dSig}):
\begin {equation}
   \int_0^{2\pi} d\phi \> d\phi' \>
         e^{-im\phi} \,\delta c_2(\v,\v')\, e^{im\phi'} \Bigg|_{v_z=v'_z=0}
   = - 8 \int_0^{2\pi} d(\Delta\phi) \>
             {\cos^2(\Delta\phi) \over | \sin(\Delta\phi) |} \,
             e^{i m \, \Delta\phi} .
\end {equation}
There is then a log singularity associated with $\Delta\phi \to 0$.
If one repeats the above for small (rather than zero) $v_z$ and $v'_z$,
one finds the dominant logarithmic behavior is
\begin {equation}
   \int_0^{2\pi} d\phi \> d\phi' \>
         e^{-im\phi} \,\delta c_2(\v,\v')\, e^{im\phi'}
   \approx - 16 \left[ \ln\left(1\over|v_z-v'_z|\right)
                             + (-)^m \ln\left(1\over|v_z+v'_z|\right)
                        \right] .
\label{eq:logs}
\end {equation}
For even $m$, if we substitute back into (\ref{eq:dSig}), treat the
$\delta$-functions as smeared over $v_z \rho \sim 1$, and ignore
non-logarithmic corrections, then the logarithms of (\ref{eq:logs})
becomes simply $\ln(1/\rho)$'s, giving
\begin {equation}
   \delta\Sigma_{m}(\rho) \approx
   {(2m+1)\over (2m)!} \left[P_m^m(0)\right]^2 {2\over\rho^2} \, \ln\rho =
   {(2m+1)!!\, (2m-1)!!\over (2m)!} \, {2\over\rho^2} \, \ln\rho .
\end {equation}
For odd $m$, we instead take the principal part terms of (\ref{eq:dSig}),
and we need to be more careful to separate the (cut-off) integral of
the joint overall scale of $v_z$ and $v'_z$ from that of the relative scale\
$\beta \equiv v_z/v'_z$:
\begin {eqnarray}
   \delta\Sigma_{m}(\rho) &\approx&
   - {(2m+1)\over (2m)!} \left[P_m^m(0)\right]^2 {1\over\pi\rho^2}
   \int_{-\infty}^{+\infty} dv_z \> dv'_z {1\over v_z} \,
        \ln \left|v_z+v'_z \over v_z - v'_z \right| \, {1\over v'_z}
\nonumber\\
   &\approx&
   - {(2m+1)\over (2m)!} \left[P_m^m(0)\right]^2 {1\over\pi\rho^2}
   \int_{-\infty}^{+\infty} {dv_z \over v_z}
   \int_{-\infty}^{+\infty} d\beta \> \ln \left|1 + \beta \over 1-\beta \right|
\nonumber\\
   &\approx&
   - {(2m+1)!!\, (2m-1)!!\over (2m)!} \, {2\over\rho^2} \, \ln\rho .
\end {eqnarray}
Our final result is then
\begin {equation}
   \Sigma_m(\rho)
   = {(2m+1)!! \, (2m-1)!!\over (2m)!} \left[
         {\pi\over 2\rho}
         + (-)^m {2\over\rho^2} \, \ln \rho
         + O\left(1\over\rho^2\right) \right] .
\label{eq:SigAsym}
\end {equation}
We do not know how to determine the non-logarithmic $O(\rho^{-2})$ piece
analytically, but fits to numerical evaluation of the
continued-fraction formula (\ref{eq:cf}) at large $\rho$ give the
leading correction to (\ref{eq:SigAsym}) to be
$\ln \rho \to \ln\rho + h_m$ with
$h_0 \simeq -0.8$, $h_1 \simeq 1.16$, and $h_2 \simeq -4.3$.



\end {document}